\documentclass[aps,pre,twocolumn,superscriptaddress,longbibliography,showpacs]{revtex4-1}

\usepackage{amsmath}
\usepackage{amssymb}
\usepackage{graphicx}
\usepackage[utf8]{inputenc}
\usepackage{hyperref}
\usepackage{color}
\usepackage[table]{xcolor}
\usepackage{hyphenat}
\begin{document}

\title{Local equilibria and state transfer of charged classical particles on a helix in an electric field}

\author{J. Plettenberg}
\email{jpletten@physnet.uni-hamburg.de}
\affiliation{Zentrum für Optische Quantentechnologien, Universität Hamburg, Luruper Chaussee 149, 22761 Hamburg, Germany}

\author{J. Stockhofe}
\affiliation{Zentrum für Optische Quantentechnologien, Universität Hamburg, Luruper Chaussee 149, 22761 Hamburg, Germany}

\author{A. V. Zampetaki}
\affiliation{Zentrum für Optische Quantentechnologien, Universität Hamburg, Luruper Chaussee 149, 22761 Hamburg, Germany}

\author{P. Schmelcher}
\email{pschmelc@physnet.uni-hamburg.de}
\affiliation{Zentrum für Optische Quantentechnologien, Universität Hamburg, Luruper Chaussee 149, 22761 Hamburg, Germany}
\affiliation{The Hamburg Centre for Ultrafast Imaging, Universität Hamburg, Luruper Chaussee 149, 22761 Hamburg, Germany}

\begin{abstract}
We explore the effects of a homogeneous external electric field on the static properties and dynamical behavior of two charged particles confined to a helix.
In contrast to the field-free setup which provides a separation of the center-of-mass and relative motion, the existence of an external force
perpendicular to the helix axis couples the center-of-mass to the relative degree of freedom leading to equilibria with a localized center of mass.
By tuning the external field various fixed points are created and/or annihilated through different bifurcation scenarios.
We provide a detailed analysis of these bifurcations based on which we demonstrate a robust state transfer  
between essentially arbitrary equilibrium configurations of the two charges that can be induced by making the external force time-dependent.
\end{abstract}


\maketitle
\section{Introduction}

One of the commonly emerging structures in nature is that of the helix. Especially in substances which are inextricably connected with life such as the DNA molecule and particular proteins, the helical structure appears to be crucial for their stability and functionality \cite{DNA1, proteins1, proteins2, proteins3}. In addition, this structure constitutes a major factor for the exhibition of peculiar physical effects in the aforementioned macromolecules such as a negative differential resistance \cite{NegativeResistance}, a proximity induced superconductivity \cite{ProximitySuperconductivity} and the existence of topological states accompanied with a quantized current \cite{Topological1}. Even more, due to their helical structure both the DNA and the helical proteins are expected to act as efficient spin-filters \cite{SpinFilters1, SpinFilters2, SpinFilters3} or as field-effect transistors \cite{ElectricGates1, ElectricGates2, ElectricGates3, ConductanceDNA} making them good candidates for the developing 
field of molecular electronics.

Meanwhile, less complex inorganic systems such as carbon nanotubes are shaped nowadays in different forms, including a helix \cite{Graphene_Helix, HelixNanotubes1, HelixNanotubes2}. Given a pre-established helical confinement, identical long-range interacting particles such as dipoles and charges can display a plethora of different intriguing phenomena \cite{HelixDipoles1,HelixDipoles2, HelixTwoBody, LongRangeInteractions,InhomogeneousHelix,WignerCrystals,NonlinearExcitations}. Even on the level of classical mechanics a helical constraint restricts the motion of the particles inducing an oscillatory effective two-body potential for repulsively interacting electric charges \cite{HelixTwoBody,LongRangeInteractions}. This potential supports a number of bound states tunable by the geometrical parameters of the confining helix. If the helix is inhomogeneous such bound states can dissociate through the scattering of the particles, or conversely be created out of the scattering continuum \cite{InhomogeneousHelix}. 
For the many-body system different crystalline configurations can be formed accompanied with non-trivial vibrational band structures (tuned  also by the geometry parameters of the helix) as a result of the complex potential landscape of the effective interactions \cite{WignerCrystals,NonlinearExcitations}.

Apart from the interaction potential also the external potential acting on the particles is crucially affected by the presence of a helical constraint. Of particular interest is the case of a constant transverse (i.e. perpendicular to the helix axis) electric field acting on helically confined charges. This is known to induce a superlattice potential for any charge carrier, whose parameters can be easily tuned by adjusting the external applied field \cite{Superlattice1,Superlattice2}. Such a field is supposed to enhance the spin-polarized transport through DNA \cite{SpinGateDNA} or helical proteins \cite{SpinGateProteins}. Even more, for non-interacting electrons in a tight-binding approach the rotation of a transverse electric field in the transverse plane leads to an adiabatic charge pumping whose current is quantized, pointing to the existence of topological  states \cite{Topological1}.

In view of the above studies the natural question arises what would be the combined effect of interactions and an external transverse electric field acting on helically confined charges. Since the two potentials (interaction and external) introduce in general different characteristic lengths of the same scale, frustration phenomena can appear leading to a rich static and dynamical behavior. A typical system presenting such a complex behavior is the well-known Frenkel-Kontorova model which consists of harmonically interacting particles on top of a sinusoidal lattice potential \cite{FrenkelKontorova}. The more involved form of both the external and the interaction potential in our case of helically confined charges makes even the two-body problem highly non-trivial, thereby motivating its study. We find that  the total potential landscape of such a system is altered significantly by tuning the external electric field, giving rise to various kinds of local equilibria which emerge through different bifurcation 
scenarios. One of the observed effects is the merging of two neighboring  potential wells  into one  for high enough external fields. Making use of this field-driven merging of the wells we can achieve through particular quench protocols a robust state transfer which involves the charges at different interparticle separations.

This paper is organized as follows. In Sec. \ref{sec:setup} we present our setup of the two charges confined to the helix in the presence of a constant electric field and discuss briefly the features of the non-interacting case. Section~\ref{sec:bif} contains the study of the static problem for different values of the external force, including a full discussion of the existing critical points and the different bifurcations. Using our knowledge of the static problem we present in Sec.~\ref{sec:transfer} some applications of our system to state transfer. Finally, Sec.~\ref{sec:con} contains our conclusions.
\section{Setup}
\label{sec:setup}
\begin{figure}[ht!]
\begin{center}
\includegraphics[width=0.45\textwidth]{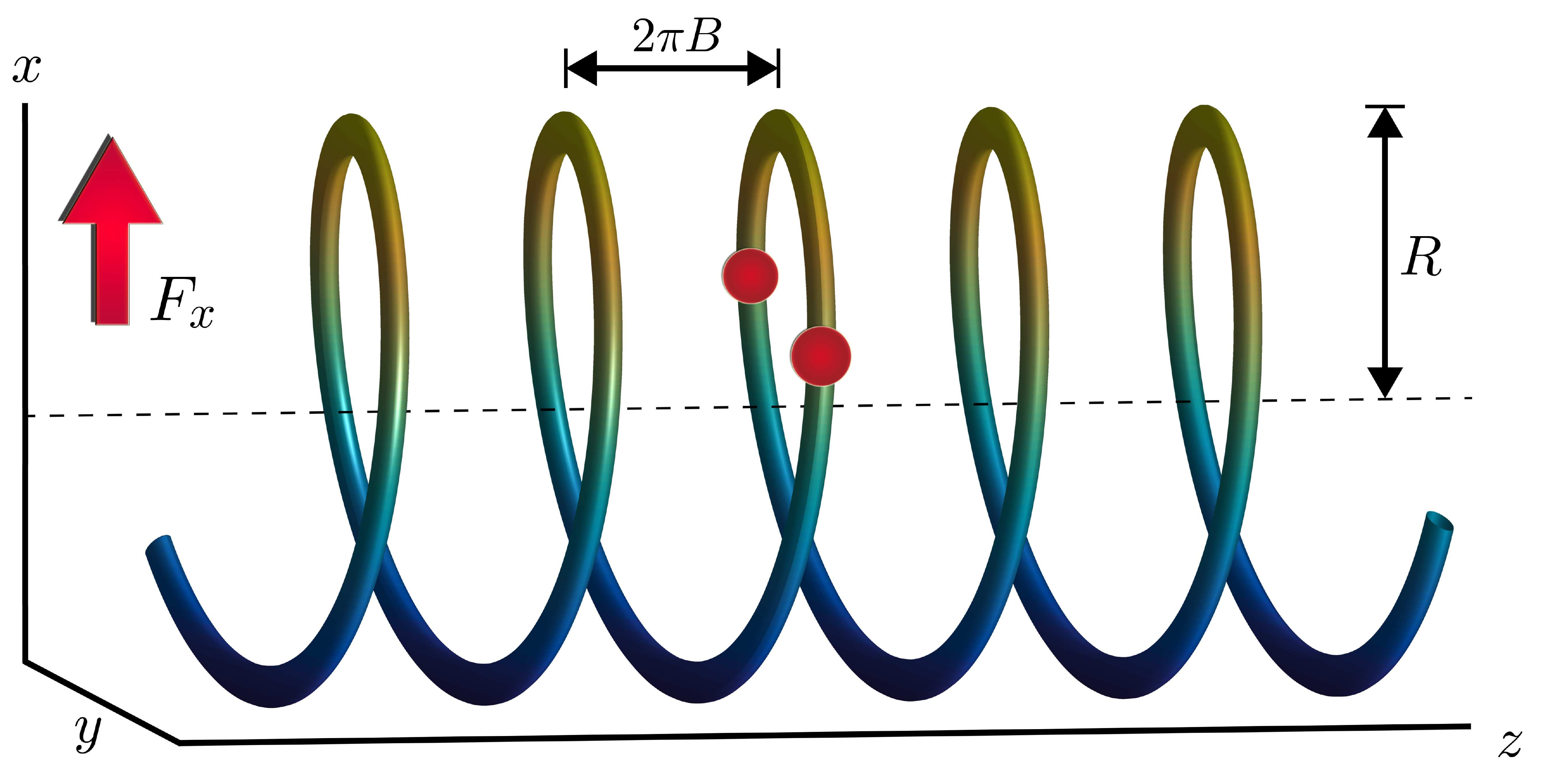}
\caption{\label{fig:setup}
(Color online) Illustration of the setup: Two identical classical charges are constrained to move on a helix curve of radius $R$ and pitch $2\pi B$. Additionally, they are subject to a homogeneous force field $\bf F$ 
(pointing along the $x$-axis in this sketch).}
\end{center}
\end{figure}

We consider a system of two identical classical particles of charge $q$ and mass $m$ each which interact via Coulomb interaction. 
The particles are constrained to move along a helix curve and exposed to a homogeneous electric field $\mathbf{E}$, thus experiencing a force $\mathbf{F}=q\mathbf{E}$. 
Due to the constraint, their position vectors $\mathbf r_i$ (where $i=1,2$ is the particle index) are uniquely determined by angular coordinates $\varphi_i \in \mathbb R$ via
\begin{equation} 
\mathbf{r}_i=\mathbf{r}(\varphi_{i})=\begin{pmatrix}
R \sin\varphi_{i},
R \cos\varphi_{i},
B \varphi_{i} \end{pmatrix},
\end{equation}
where $R$ denotes the helix radius and $2\pi B$ the pitch (see Fig.~\ref{fig:setup}).

Then the classical Lagrangian including the kinetic energy, the external potential induced by the force $\mathbf F$ and the Coulomb repulsion (with coupling constant $g>0$) 
is given by
\begin{align}
&L( \varphi_{1} , \varphi_2, \dot{\varphi}_{1}, \dot{\varphi}_{2} ) \nonumber\\
&=  \sum_{i=1}^{2} \left(\frac{m}{2} \vert \partial_{\varphi_{i}} \textbf{r}(\varphi_{i}) \vert^{2}
 \dot{\varphi}_{i}^{2}
+ \textbf{F} \cdot \textbf{r}(\varphi_{i}) \right)
- \frac{g}{\vert \textbf{r}(\varphi_{1}) - \textbf{r}(\varphi_{2}) \vert} \nonumber\\
&=  \sum_{i=1}^{2}
\left( \frac{m}{2} (R^{2}+B^{2})\dot{\varphi}_{i}^{2}+F_{x}R \sin \varphi_{i}+F_{y}R \cos \varphi_{i} +F_{z} B \varphi_{i} \right)
\nonumber \\
&\qquad -\frac{g}{\sqrt{2R^{2} \left[1-\cos (\varphi_{1}-\varphi_{2})\right]
+ B^{2} (\varphi_{1}-\varphi_{2})^{2}}}.
\label{eq:Lagrangian}
\end{align}
Note the appearance of a geometry factor $R^2+B^2$ in the kinetic term here, such that when switching to arclength coordinates $s_i = \varphi_i \sqrt{R^2+B^2}$ the canonical form $\frac{m}{2}\dot s_i^2$ would be restored.
The Coulomb repulsion potential is governed by the Euclidean interparticle distance in the full three-dimensional space, which gives rise to an intricate non-monotonic effective potential in the one-dimensional angular coordinates.
By virtue of this geometry-induced deformation of the interaction potential metastable bound states can be formed for charged particles constrained to a helix, despite the underlying Coulomb force being purely repulsive \cite{LongRangeInteractions}.
In a similar fashion, the presence of the constraint turns the homogeneous three-dimensional force field into a nontrivial potential landscape when viewed in the angular coordinates.
At a given position along the helix curve, a particle only feels the component of the force that is locally tangential to the curve which gives rise to an oscillatory structure in the effective one-dimensional potential.

Let us start by briefly discussing the case of a single particle on the helix. From Eq. (\ref{eq:Lagrangian}), the effective potential reads
\begin{align}
\label{eq:1ParticleExternalPotential}
V(\varphi_1)=-F_{x}R \sin \varphi_1-F_{z} B \varphi_1,
\end{align}
where we have set $F_{y}=0$ and consider $F_x > 0$ without loss of generality (rotating the coordinate axes and shifting the origin).
Clearly, $F_{z}$ (the force component parallel to the helix axis) induces a linear decrease of the potential energy, thus causing constant acceleration. 
In contrast, $F_{x}$  (the force component perpendicular to the helix axis), gives rise to a spatially oscillating contribution to the effective potential.
By itself, this contribution has minima at $\varphi_{1}=\frac{\pi}{2} + 2 \pi k$, $k \in \mathbb{Z}$, i.e. when the particle is maximally pushed to the far end of the helix by the perpendicular force,
and correspondingly maxima at $\varphi_{1}=-\frac{\pi}{2} + 2 \pi k$, $k \in \mathbb{Z}$.
Whether there are fixed points in the presence of both $F_{x}$ and $F_{z}$ depends on their ratio as well as on the geometry of the helix. Specifically, if
\begin{equation}
\label{eq:SingleParticle_ConditionWheteherWellsExist}
\frac{R}{B} \frac{F_{x}}{\vert F_{z} \vert} \geq 1,
\end{equation}
there is a spatially periodic sequence of maxima and minima, while otherwise no fixed points exist and there is no bound motion.

\section{Critical points and bifurcations}
\label{sec:bif}
Let us now turn to the full two-body problem.
As can be seen from Eq.~(\ref{eq:Lagrangian}), the deformed interaction potential depends on the angle difference $\varphi_1-\varphi_2$ only,
which in the absence of the external force leads to a separation of the center-of-mass (COM) and relative motion \cite{LongRangeInteractions,InhomogeneousHelix}.
We thus transform to the relative coordinate $\varphi=\varphi_{1}-\varphi_{2}$ and the COM coordinate $\Phi=\frac{\varphi_{1}+\varphi_{2}}{2}$,
such that the Lagrangian turns into
\begin{align}
&L(\varphi,\Phi,\dot{\varphi},\dot{\Phi})
=  m(R^2+B^2) \dot{\Phi}^{2} + \frac{m}{4}(R^2+B^2)\dot{\varphi}^{2} \nonumber \\
&-\frac{g}{\sqrt{2 R^{2} ( 1-\cos\varphi) 
+ B^{2} \varphi^{2} }}
+2 F_{x}R \sin\Phi \cos \frac{\varphi}{2} 
+2F_{z} B \Phi,
\label{eq:Lagrangiancom}
\end{align}
where again we have restricted ourselves to $F_y=0$ and $F_x>0$.
It can be seen that the $F_z$ component does not break the COM separation and only induces a constant acceleration of $\Phi$.
In contrast, $F_x$ couples $\varphi$ and $\Phi$ and impedes the decoupling. We are mostly interested in the interplay of this $F_x$-induced term and the geometrically modified interaction and 
will therefore focus on $F_{z}=0$ in the following.  

Let us first switch to dimensionless units according to
\begin{align}
\tilde{x}=\frac{x}{R}, \hspace{4mm}
\tilde{t}=t \sqrt{\frac{g}{m R^{3}}}, \hspace{4mm}
\tilde{L}=\frac{L R}{g}, \hspace{4mm}
\tilde{F}_{x,y,z}=F_{x,y,z} \frac{R^2}{g},
\end{align}
where $\tilde x$ and $\tilde t$ denote rescaled length and time variables, $\tilde L$ the rescaled Lagrangian and $\tilde{F}_{x,y,z}$ the rescaled force components. The rescaled pitch parameter is given by $b=\frac{B}{R}$.
Effectively, this amounts to setting $m=g=R=1$.
We will omit the tildes in the following. In dimensionless units, the potential term in the Lagrangian of Eq.~(\ref{eq:Lagrangiancom}) reads (with $F_z=0$)
\begin{align}
\label{eq:VCoulombRescaled}
&V(\varphi,\Phi)=\frac{1}{\sqrt{2 - 2\cos\varphi+ b^{2} \varphi^{2} }}
-2F_{x} \sin\Phi\cos \frac{\varphi}{2}.
\end{align}

\begin{figure}[ht!]
\begin{center}
\includegraphics[width=0.4\textwidth]{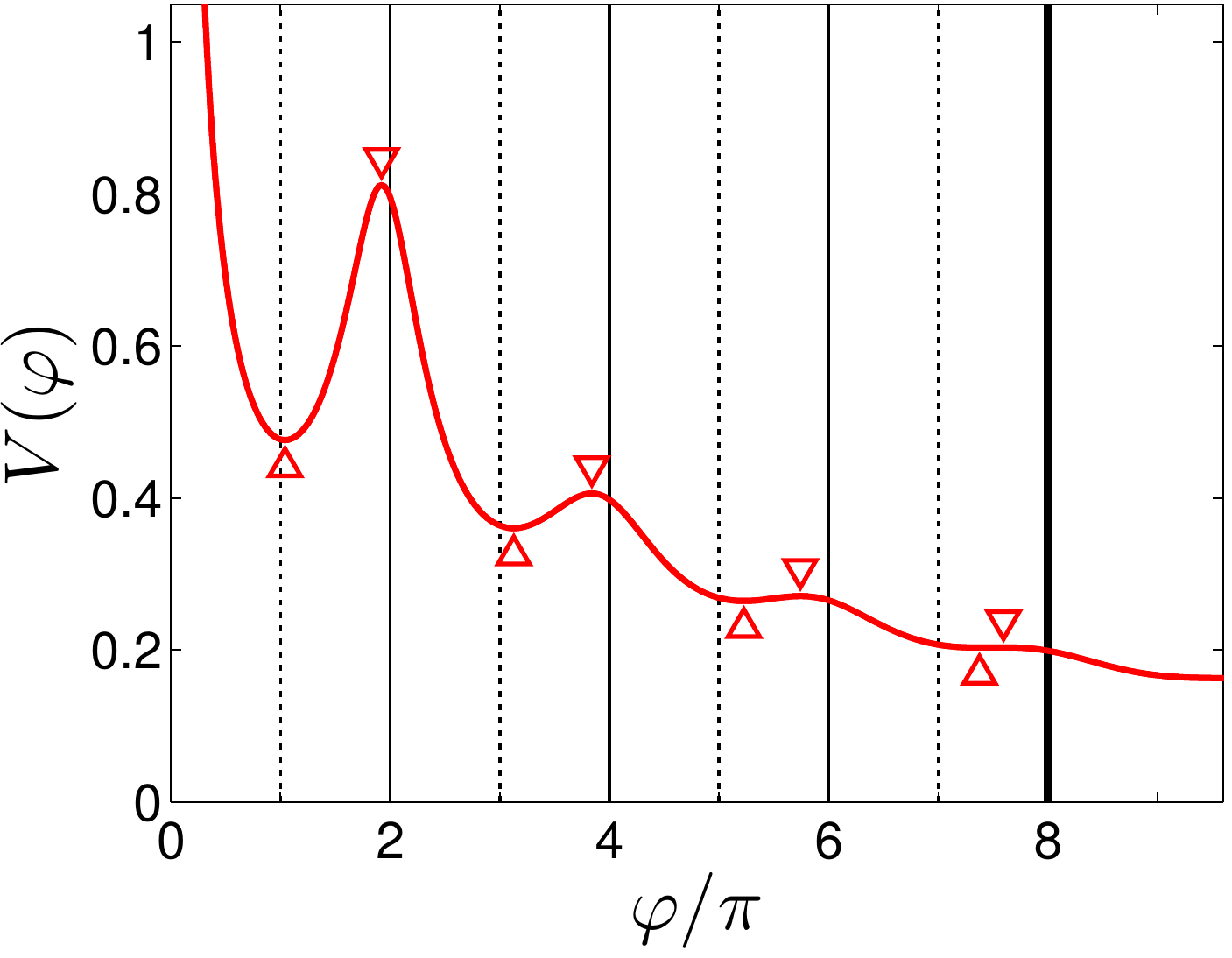}
\caption{\label{fig:VCoulomb1}
(Color online) Two-body interaction potential as a function of the relative coordinate $\varphi$ for $b=0.2$. Triangular markers pointing upwards (downwards) indicate the locations of local minima (maxima).
Beyond a critical $\varphi=2\pi r_c$, indicated by the thick vertical line, no further maxima or minima exist ($r_c=4$ here).}
\end{center}
\end{figure}

The rest of this section is devoted to a discussion of the stationary points of this effectively two-dimensional potential landscape, 
which correspond to the fixed points of the two-body problem, and their bifurcations as the external force $F_x$ is varied.
Let us therefore start by recapitulating the fully force-free case of $F_{x}=0$ \cite{LongRangeInteractions} in which the COM moves freely. 
The interaction potential is dominated by the oscillatory cosine term for $b \varphi \ll 1$, while in the opposite limit of $b \varphi \gg 1$ the asymptotic repulsive Coulomb tail is recovered,
as can be clearly seen in the example shown in Fig.~\ref{fig:VCoulomb1}.
Depending on $b$ (i.e. on the pitch-to-radius ratio of the helix), the oscillatory contribution to the interaction potential can induce a finite number of metastable bound states with respect to $\varphi$.
For $b=0.2$ as in Fig.~\ref{fig:VCoulomb1}, there are four such minima (and four maxima). Generally, $V(\varphi)$ possesses a minimum-maximum pair in every interval $[2\pi r - \pi,2\pi r], r \in \mathbb N,$ up to some critical integer $r_c$ which depends on $b$ ($r_c=4$ in the example shown in the figure). We will refer to the integer $r$ as the ``order'' of the associated pair of critical points. 
With increasing order $r$, the potential wells around the minima become shallower due to the impact of the Coulomb repulsion at larger $\varphi$.
The integer $r_c$ that counts the total number of minima (and maxima) decreases with increasing $b$ and eventually,
for $b \gtrsim 0.466$, there are no extrema anymore and $r_c =0$.

\begin{figure}[ht!]		
\includegraphics[width=0.5\textwidth]{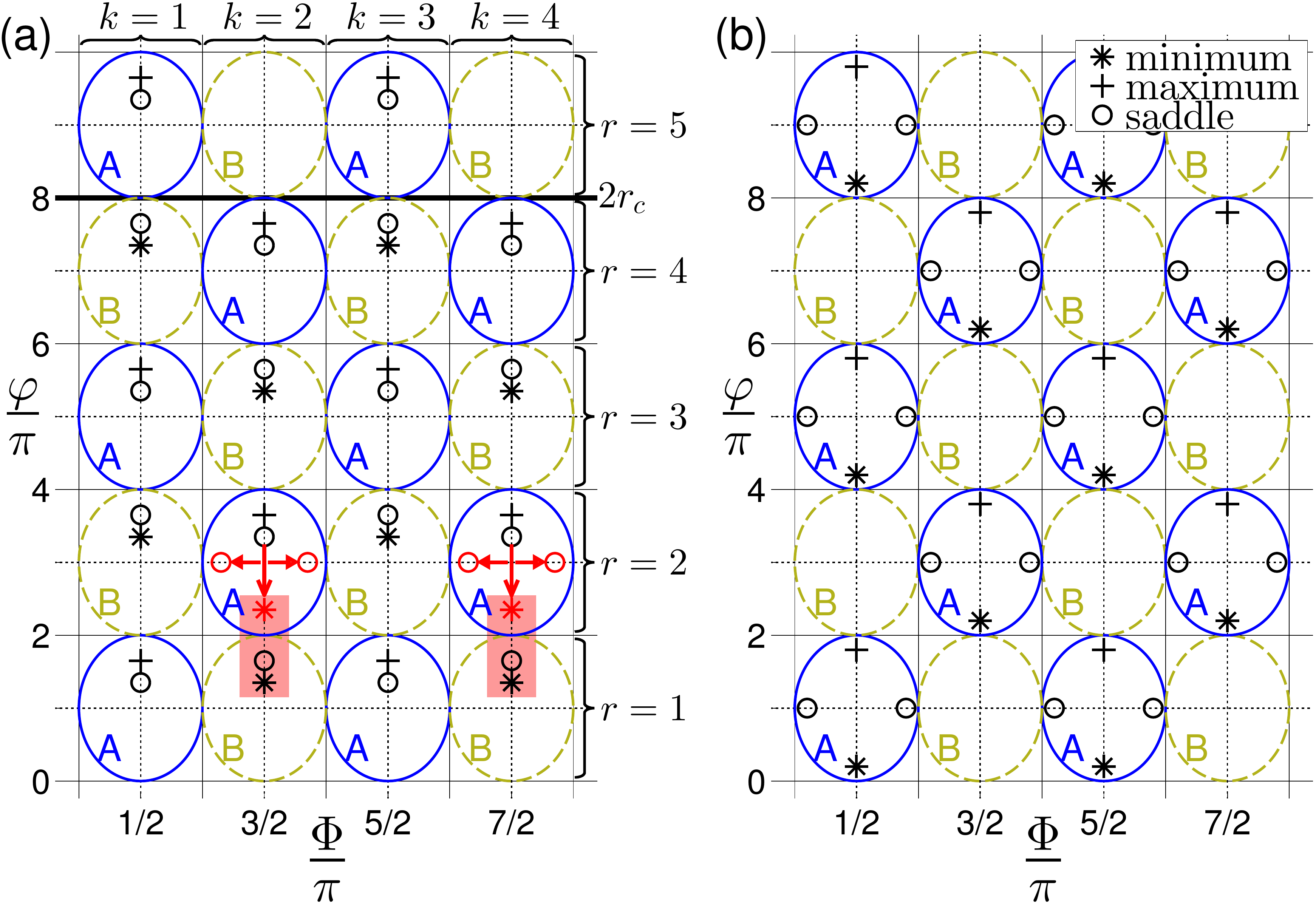}
\caption{\label{fig:EquilibriaLandscape}
(Color online) Schematic view of the fixed-point landscape (not to scale) for (a) small values of $F_x$ and (b) large values of $F_x$. Fixed points are indicated by markers, where circles stand for saddles, 
asterisks for minima and pluses for maxima, respectively. Different types of configuration space cells (A and B, see text) are marked by solid blue and dashed green ellipses, respectively.
In the type-A cells of $r=2$, the fixed points emerging in the subcritical pitchfork bifurcation are indicated in red. This bifurcation results in two close-lying minima at the same $\Phi$ with $r=1$ and $r=2$, respectively, separated by a saddle, as
highlighted by the red shaded areas. Protocols for controlled transfer between these minima are discussed in Sec.~\ref{sec:transfer}.}
\end{figure}

For $F_x=0$ the above discussion of the critical points applies with respect to the relative coordinate $\varphi$, while the COM is free such that all fixed points of the full two-dimensional potential $V(\varphi,\Phi)$ are neutrally stable
with respect to the $\Phi$-direction.
Switching on a small $F_x>0$, the fixed points immediately localize with respect to the COM coordinate $\Phi$ due to the second term in Eq.~(\ref{eq:VCoulombRescaled}).
Specifically, fixed points exist at the maxima and minima of $\sin \Phi$, i.e. at $\Phi=\frac{\pi}{2}+k\pi,$ $k \in \mathbb{Z}$.
The stability of the two-dimensional fixed points with respect to the $\Phi$-direction is governed by the product $\sin\Phi \cos \frac{\varphi}{2}$ and thus depends both on $k$ and on the order index $r$ with respect to the relative coordinate.
This results in a fixed point landscape as schematically shown in Fig.~\ref{fig:EquilibriaLandscape}(a).
The full configuration space can be divided into cells of length $\pi$ in the $\Phi$-direction and length $2\pi$ in the $\varphi$-direction, where each cell (and the fixed points it contains) can be labeled by the index $k$ and the order $r$ introduced above. 
Fixed points with $k+r$ even are unstable in the $\Phi$-direction, thus each minimum-maximum pair with respect to $\varphi$ turns into a two-dimensional saddle-maximum pair (cells of type A marked by blue solid ellipses in Fig.~\ref{fig:EquilibriaLandscape}) as an infinitesimal $F_x$ is switched on.
In contrast, for $k+r$ odd the fixed points are stable with respect to $\Phi$ and thus any minimum-maximum pair with respect to $\varphi$ results in a two-dimensional minimum-saddle pair (cells of type B marked by green dashed ellipses in Fig.~\ref{fig:EquilibriaLandscape}). While cells with $r > r_c$ contain no fixed points in the force-free limit $F_x=0$, this changes already at extremely small values of $F_x$: For each type-A cell with $r>r_c$ a saddle-maximum pair emerges in a saddle-node bifurcation, while in contrast type-B cells with $r>r_c$ remain empty of fixed points for any $F_x$.
The fate of the fixed points when $F_x$ is further increased is radically different for the type-A cells and the type-B cells: 
In the A cells, there is a pitchfork bifurcation scenario leading to a quartet of two saddles, a maximum and a minimum at large $F_x$, 
while in the B cells all fixed points are annihilated in saddle-node events as $F_x$ is increased. 
Ultimately, at large values $F_x$, this leads to a checkerboard pattern of fixed points as illustrated in Fig.~\ref{fig:EquilibriaLandscape}(b), which is governed by the $F_x$-term in Eq.~(\ref{eq:VCoulombRescaled}). 
In the following subsections, we provide a more detailed analysis of the bifurcations in the two types of configuration space cells.

\subsection{Subcritical pitchfork bifurcations}
Let us first focus on the fixed points in the cells of type A, with $k+r$ even, marked by blue ellipses in Fig.~\ref{fig:EquilibriaLandscape}. For small values of $F_x$, these cells contain a saddle and a maximum each. For $r\leq r_c$, this is true even in the force-free limit, and for $r>r_c$ such a pair forms at very small values of $F_x$ (see below for a more quantitative estimate). 
Let us note first that all type-A configuration space cells of the same order $r$ are fully equivalent, since the potential $V(\varphi,\Phi)$ is invariant under $\Phi \rightarrow \Phi+2\pi$. In contrast, there is no periodicity in $\varphi$ due to the Coulomb term.

With increasing $F_{x}$, the relative distances between the particles in the equilibrium configurations change while the COM coordinate remains localized at $\Phi=\frac{\pi}{2}+\pi k$. Thus, in the two-dimensional landscape as sketched in Fig.~\ref{fig:EquilibriaLandscape} the fixed points in the type-A cells move along the $\varphi$-axis. Specifically, in each of these cells the initial saddle moves towards smaller $\varphi$ and eventually crosses $\varphi=(2r-1)\pi$, 
i.e. it moves into the lower half of the cell. This can be seen to be accompanied by a stability change: The initial saddle turns into a minimum. In parallel, two additional saddles emerge which subsequently remain at $\varphi=(2r-1)\pi$ but separate along the $\Phi$-axis (as indicated in red in the A cells of order $r=2$ in Fig.~\ref{fig:EquilibriaLandscape}).

\begin{figure}[ht!]		
\centering
\includegraphics[height=0.24\textwidth]{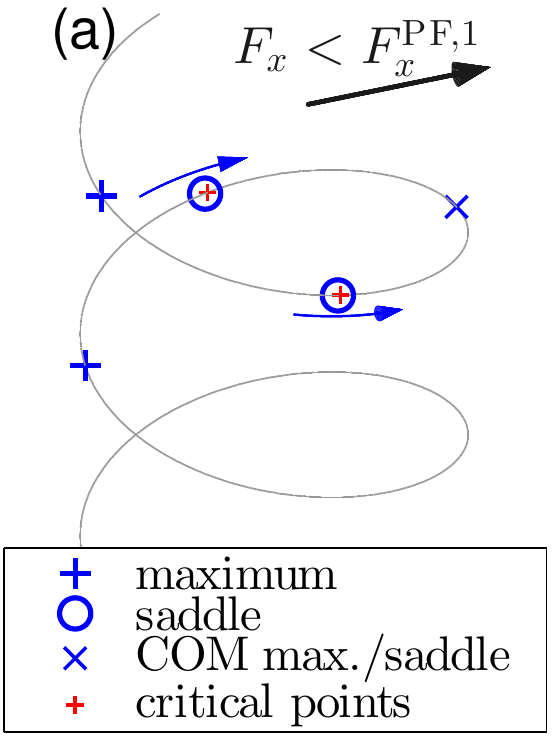} \hspace{6mm}
\includegraphics[height=0.24\textwidth]{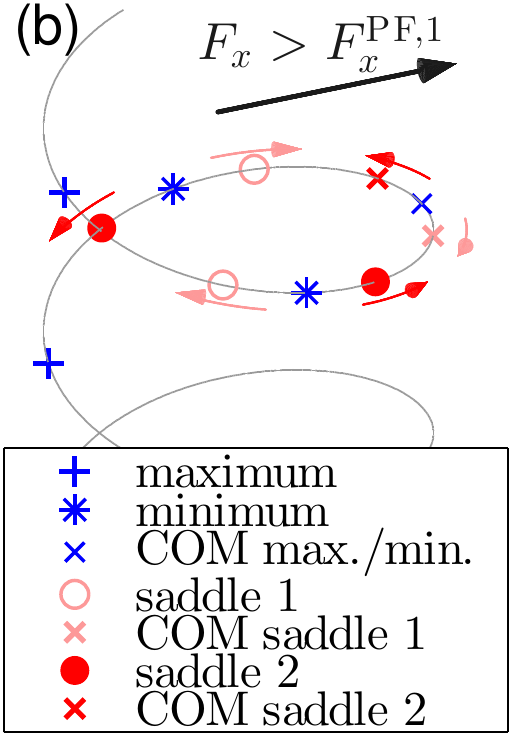}
\includegraphics[width=0.4\textwidth]{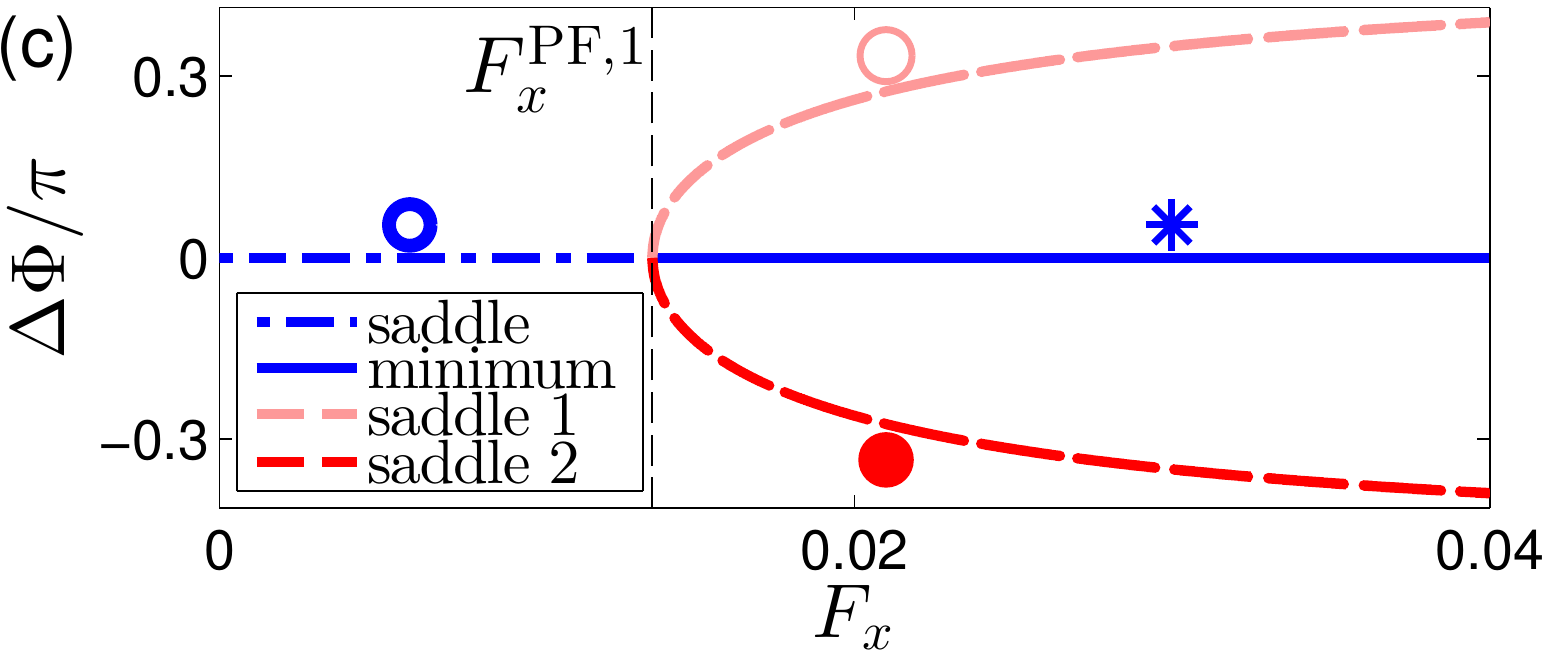}
\includegraphics[width=0.23\textwidth]{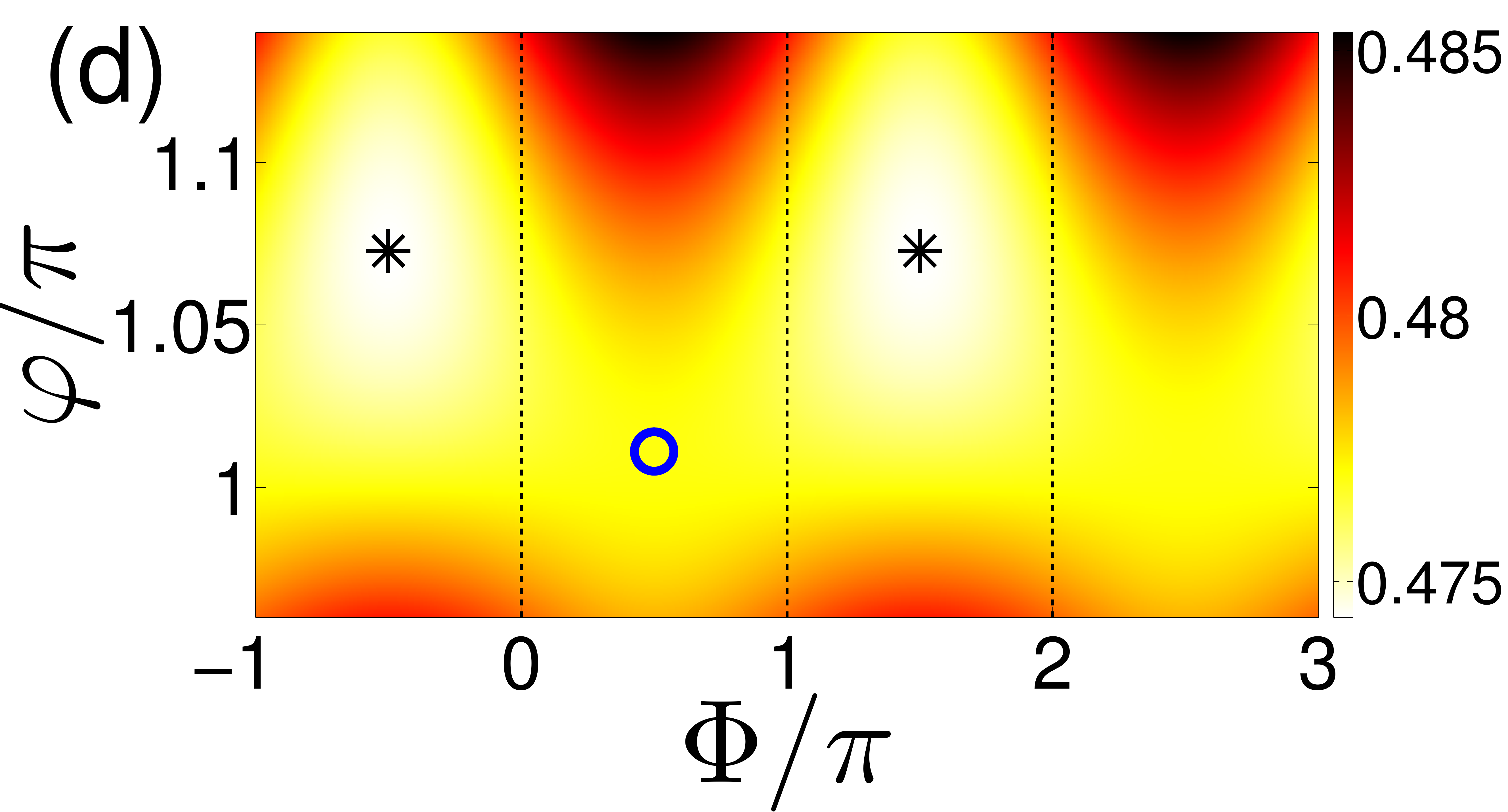}
\includegraphics[width=0.23\textwidth]{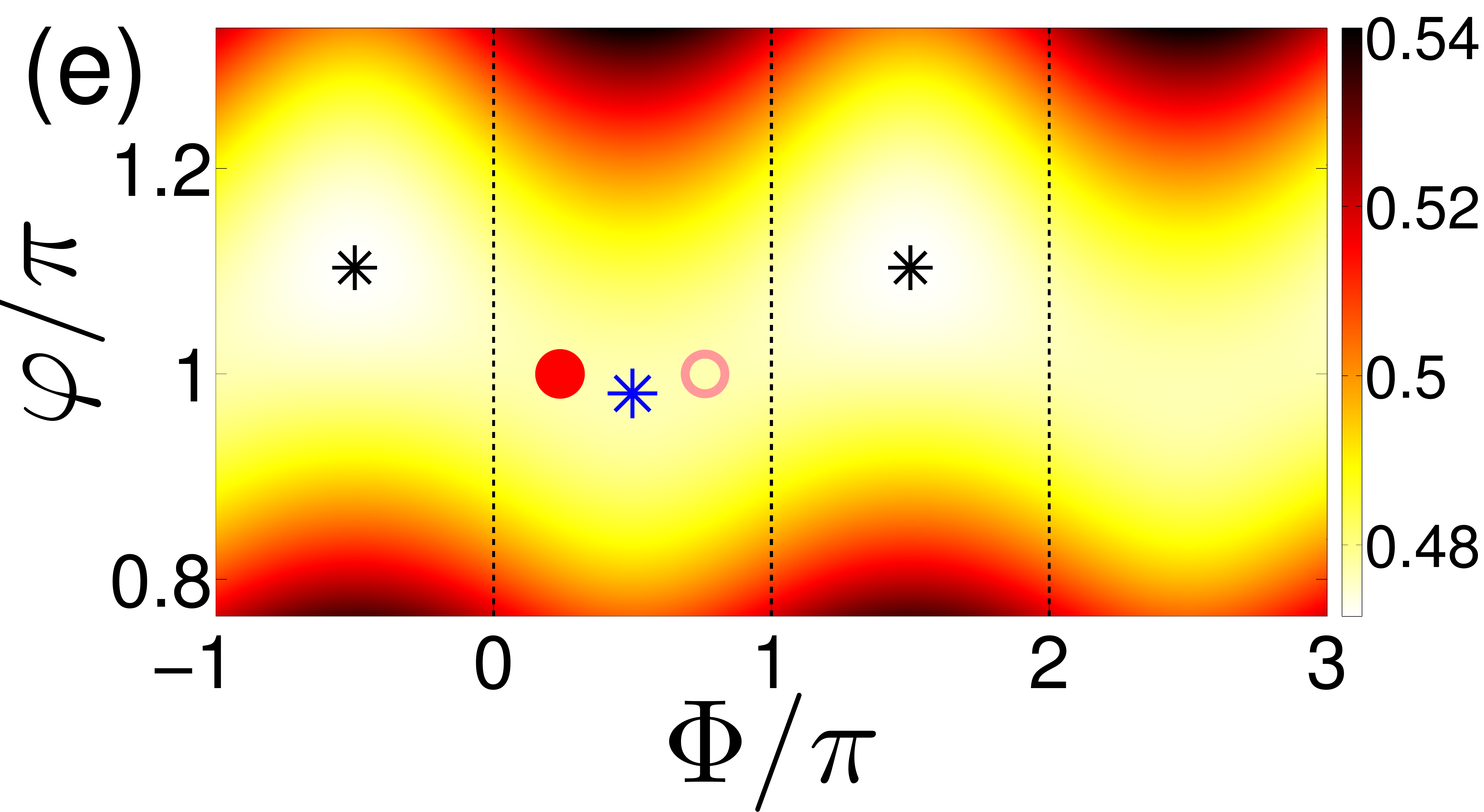}
\caption{\label{fig:PitchforkFigures} 
(Color online) Subcritical pitchfork bifurcation in configuration cells of type A, $r=1$, $b=0.2$. (a) Equilibria below the critical force, $F_x < F_x^{\text{PF,}1}$. Plus and circle markers indicate the particle positions in the maximum and saddle configuration, respectively, arrows show the directions into which the latter move with increasing $F_x$. (b) Equilibria above the critical force, $F_x > F_x^{\text{PF,}1}$. The prior saddle has turned into a minimum and new saddles have emerged whose center-of-mass coordinates drift with increasing $F_x$. (c) Bifurcation diagram showing the displacement of the COM coordinate $\Delta \Phi = \Phi-(\frac{\pi}{2}+\pi k)$ in the corresponding equilibrium configurations as in (a,b) versus $F_x$.
(d,e) Color-encoded profiles of the potential landscape $V(\varphi,\Phi)$ for $F_x < F_x^{\text{PF,}1}$ and $F_x > F_x^{\text{PF,}1}$, respectively, with locations of the fixed points indicated by the same markers as in (c). Dashed lines denote the edges of the configuration space cells. The minima in the neighboring type-B cells are also shown as black asterisk markers.}
\end{figure}
Figs.~\ref{fig:PitchforkFigures}(a,b) illustrate the three-dimensional configurations of the charges on the helix that correspond to these fixed points (for $r=1$ and $b=0.2$). 
The maximum configuration (plus markers) is characterized by the charges sitting roughly one winding apart on the side of the helix that faces towards the force, thus maximizing the field-induced potential energy. This configuration is only weakly modulated by increasing $F_x$.
In contrast, in the initial saddle configuration (circle markers in Fig.~\ref{fig:PitchforkFigures}(a)) the charges are located a bit more than half a winding apart and pushed towards the same edge of the helix by the increasing force,
which reduces their angular distance. When the distance drops below a half-winding, the configuration becomes fully stable (also with respect to translating its center of mass). This is accompanied by the emergence of two new equilibria in which the charges are separated by exactly half a winding, but with a displaced center of mass. Inspecting the COM coordinates of the fixed points vs. $F_x$ reveals the typical subcritical pitchfork bifurcation form, in which the symmetric equilibrium (with respect to reflection about the initial $\Phi=\frac{\pi}{2}+\pi k$) gains stability while two unstable symmetry-broken solutions emerge, see Fig.~\ref{fig:PitchforkFigures}(c).
Panels (d,e) of Fig.~\ref{fig:PitchforkFigures} show the corresponding deformation of the two-dimensional potential landscape $V(\varphi,\Phi)$.

\begin{figure}[ht!]
\centering
\includegraphics[width=0.49\textwidth]{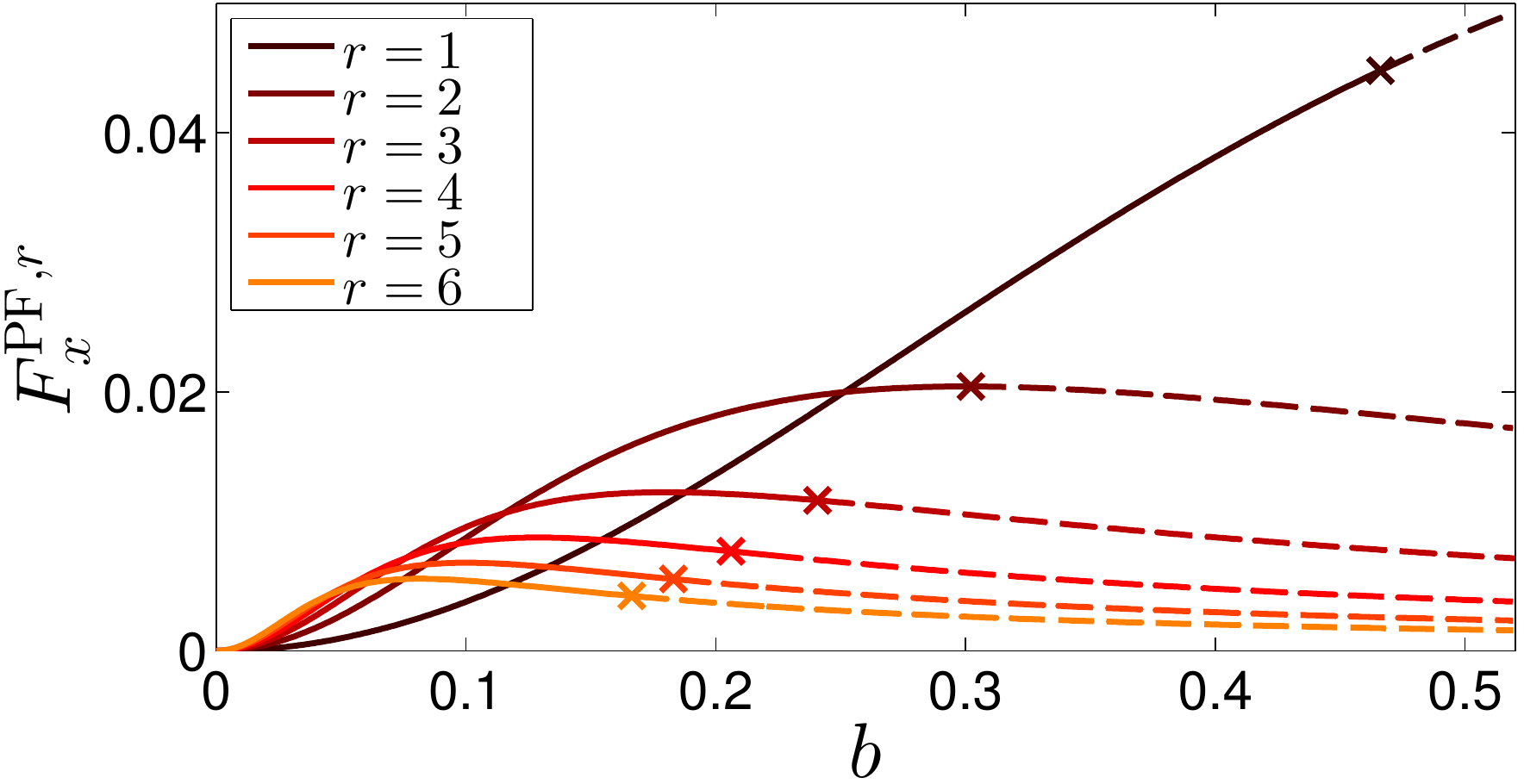}
\caption{\label{fig:F_zigzag} 
(Color online) Critical force at which the subcritical pitchfork bifurcation takes place in the type-A configuration space cells as a function of $b$ and the order $r$ of the cell.
Note the non-monotonic dependence both on $r$ and $b$. 
For values of $b$ for which the lines are dashed (beyond the cross markers), the involved fixed points do not exist in the force-free limit but are created in a saddle-node bifurcation at some $F_x < F_x^{\text{PF,}r}$.}
\end{figure}

Quantitatively, the critical force at which the subcritical pitchfork bifurcation occurs in the type-A cells of order $r$ is found to be
\begin{align} \label{eq:F_Pitchfork}
F_{x}^{\text{PF,}r}=\frac{b^{2}\pi (2r-1)}{[4+b^{2}\pi^{2}(2r-1)^{2}]^{3/2}},
\end{align}
which at the same time provides an upper bound on the critical force at which the saddle-maximum pair emerges in the initially empty cells with $r>r_c$.
While for large $r$ the critical force $F_{x}^{\text{PF,}r}$ always goes to zero, its overall dependence on $r$ is an intricate one and is also controlled by $b$ in a non-monotonic way, see Fig.~\ref{fig:F_zigzag}.


\subsection{Saddle-node bifurcations}
Let us now turn to the configuration space cells of type B, with $k+r$ odd (indicated by green ellipses in Fig. \ref{fig:EquilibriaLandscape}).
Again, translational invariance in the $\Phi$-direction ensures that such type-B cells of different $k$ but identical $r$ are equivalent. 
For $r>r_c$, the type-B cells are empty of fixed points in the force-free limit and remain empty for any $F_x$. 
For $r\leq r_c$, there is a minimum and a saddle in each such cell for infinitesimal $F_x>0$.
Increasing $F_x$ further, these approach each other along the $\varphi$-direction and eventually undergo saddle-node annihilations at a critical force $F_x^{\text{SN,}r}$, such that at large $F_x$ all type-B fixed points have disappeared.

\begin{figure}[ht!]
%
\centering
\includegraphics[height=0.23\textwidth]{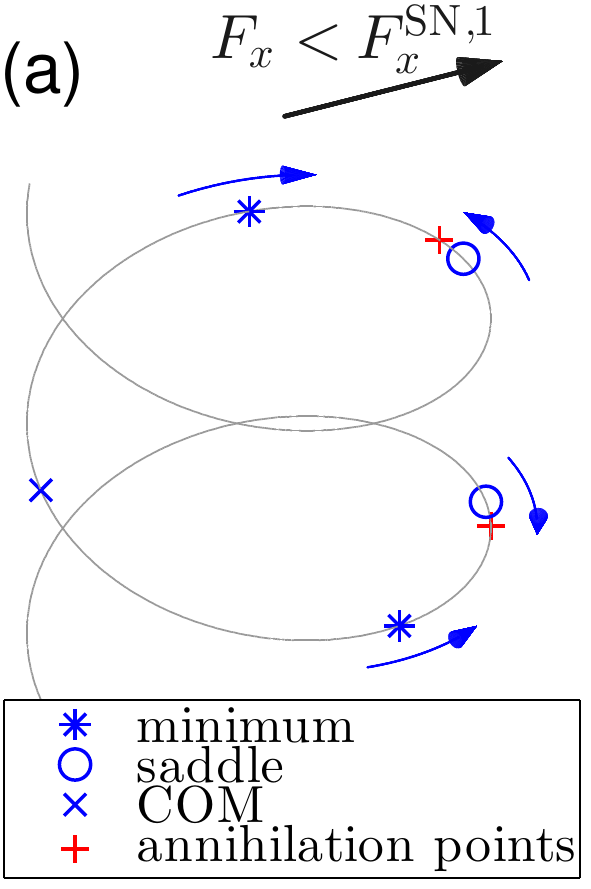} \hspace{8mm}
\includegraphics[height=0.23\textwidth]{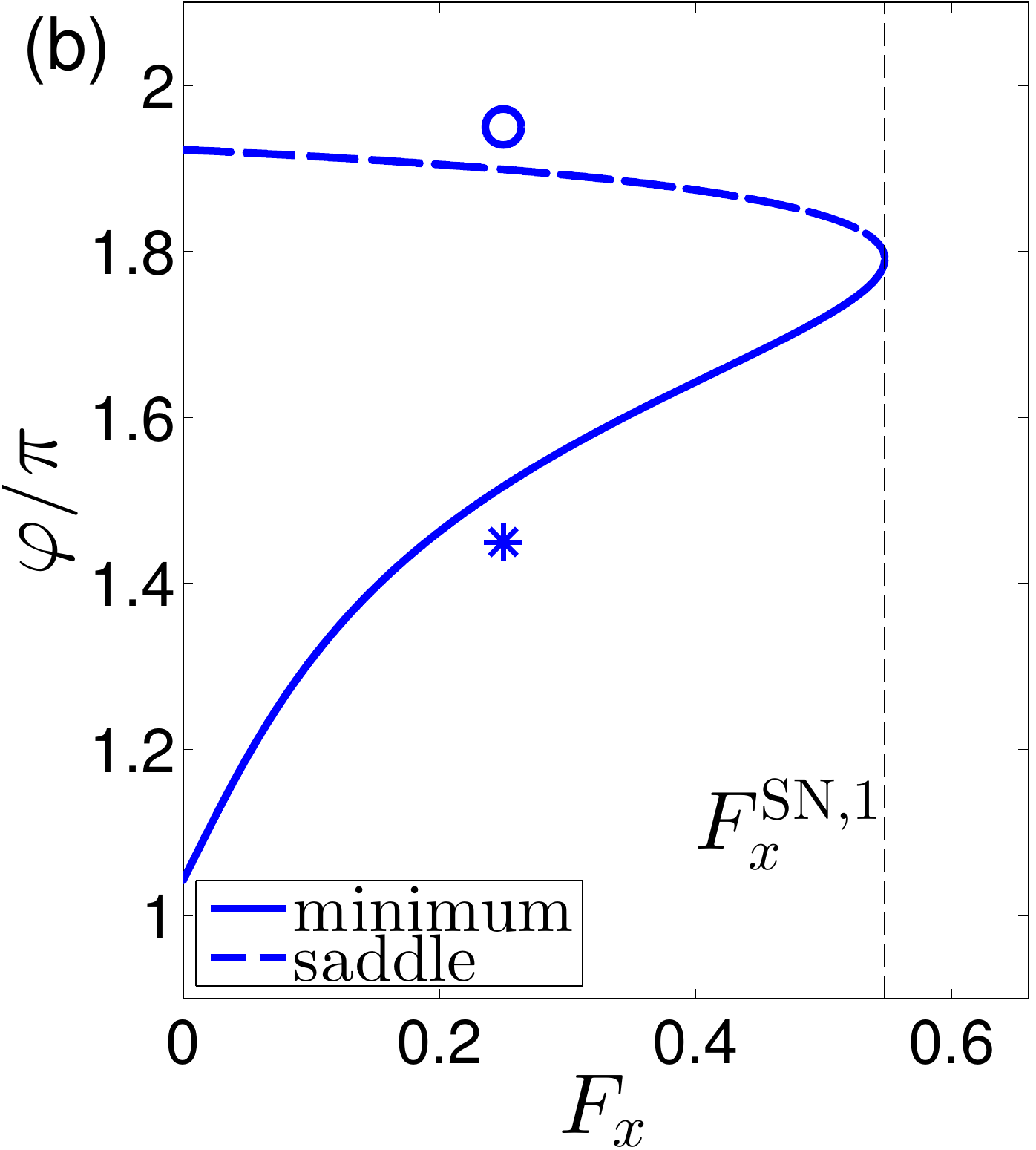}
\includegraphics[width=0.23\textwidth]{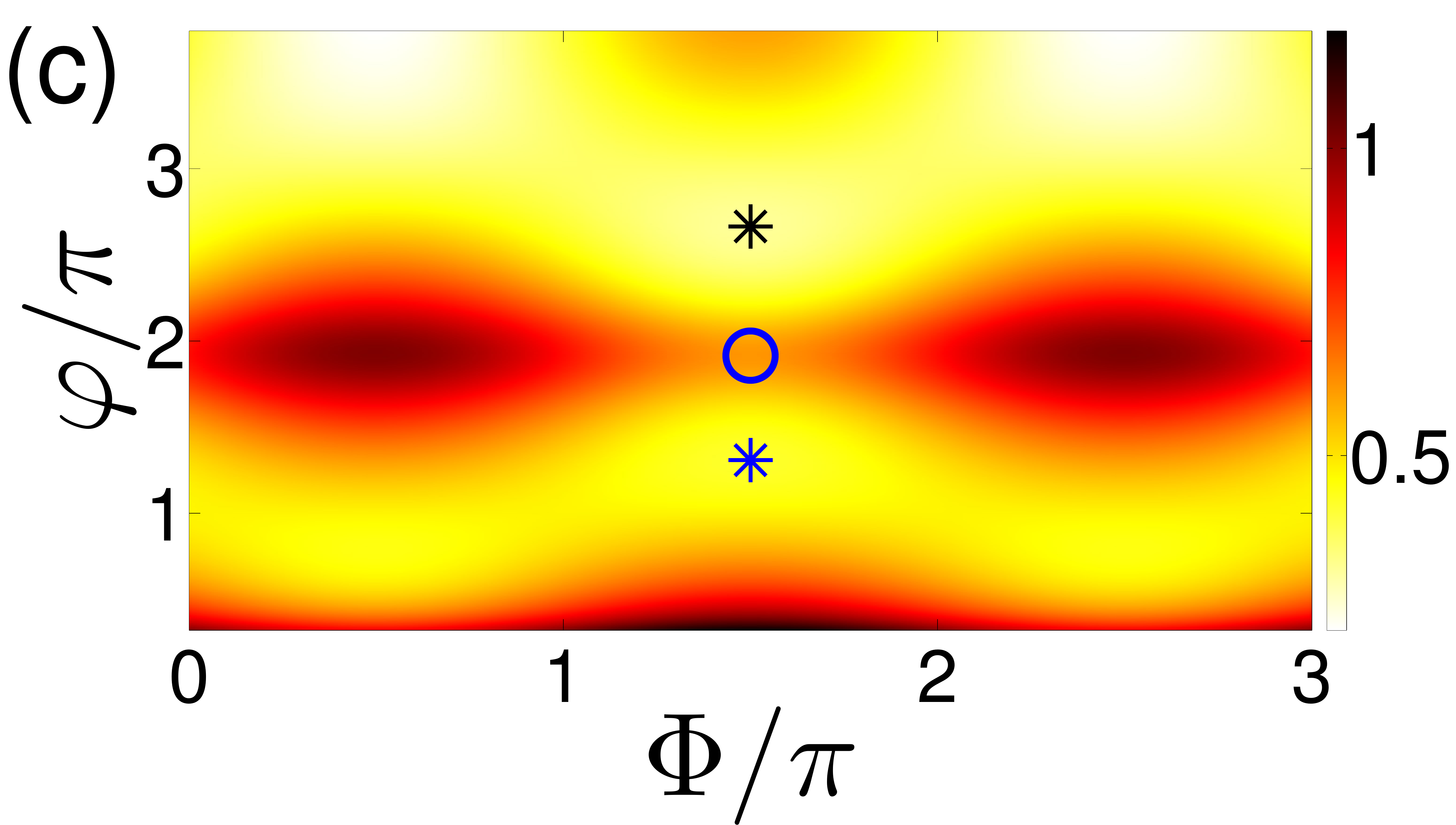}
\includegraphics[width=0.23\textwidth]{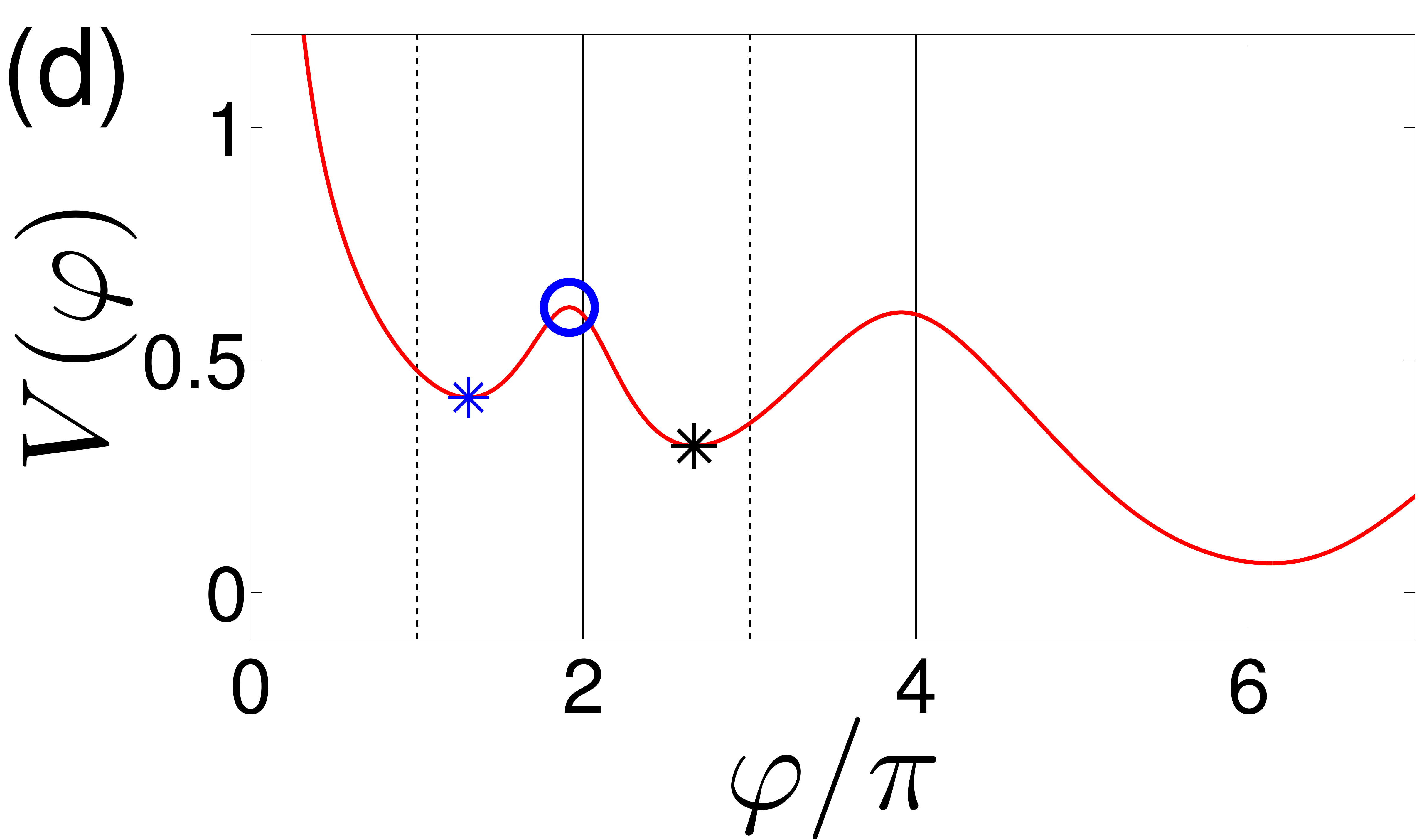}
\includegraphics[width=0.23\textwidth]{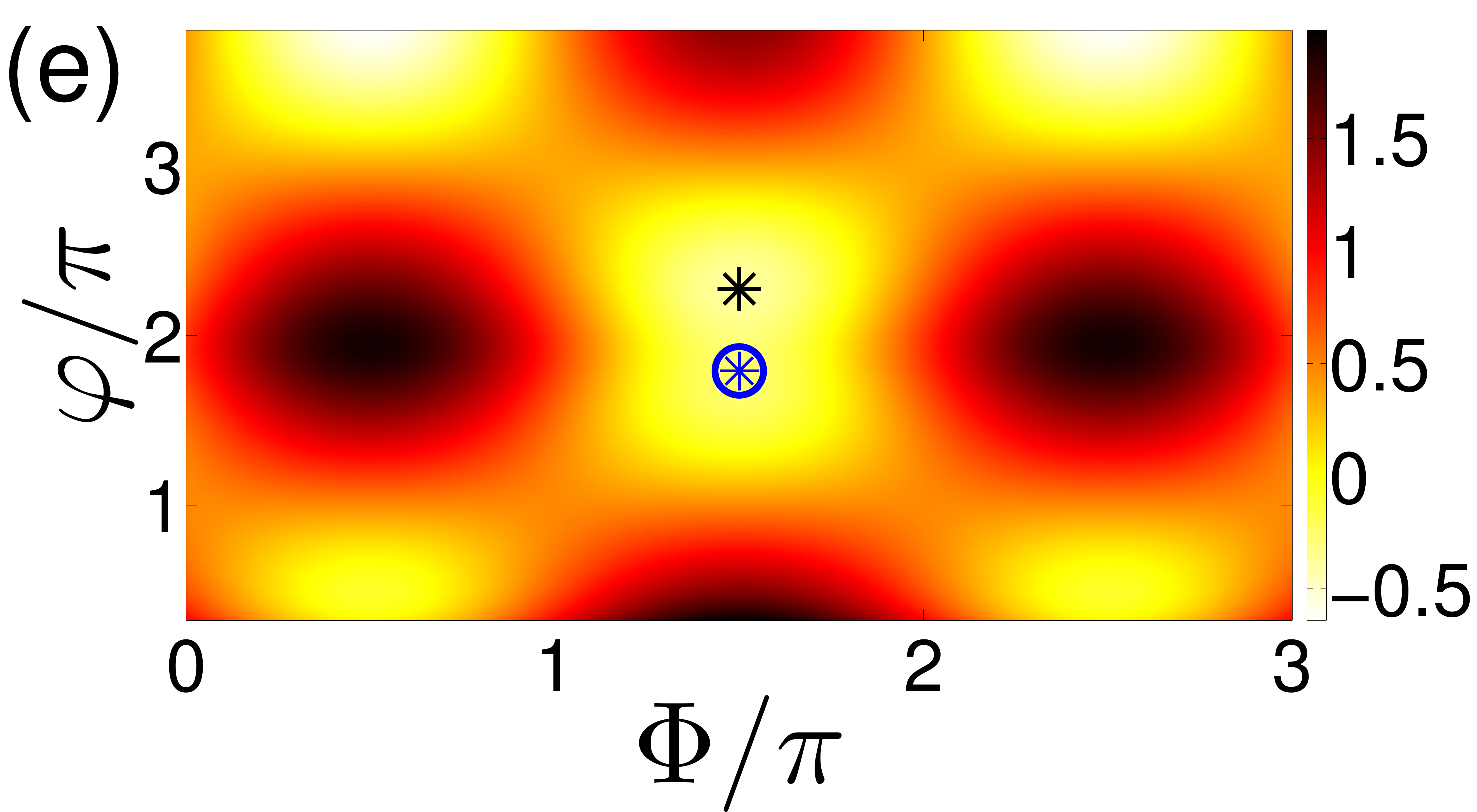}
\includegraphics[width=0.23\textwidth]{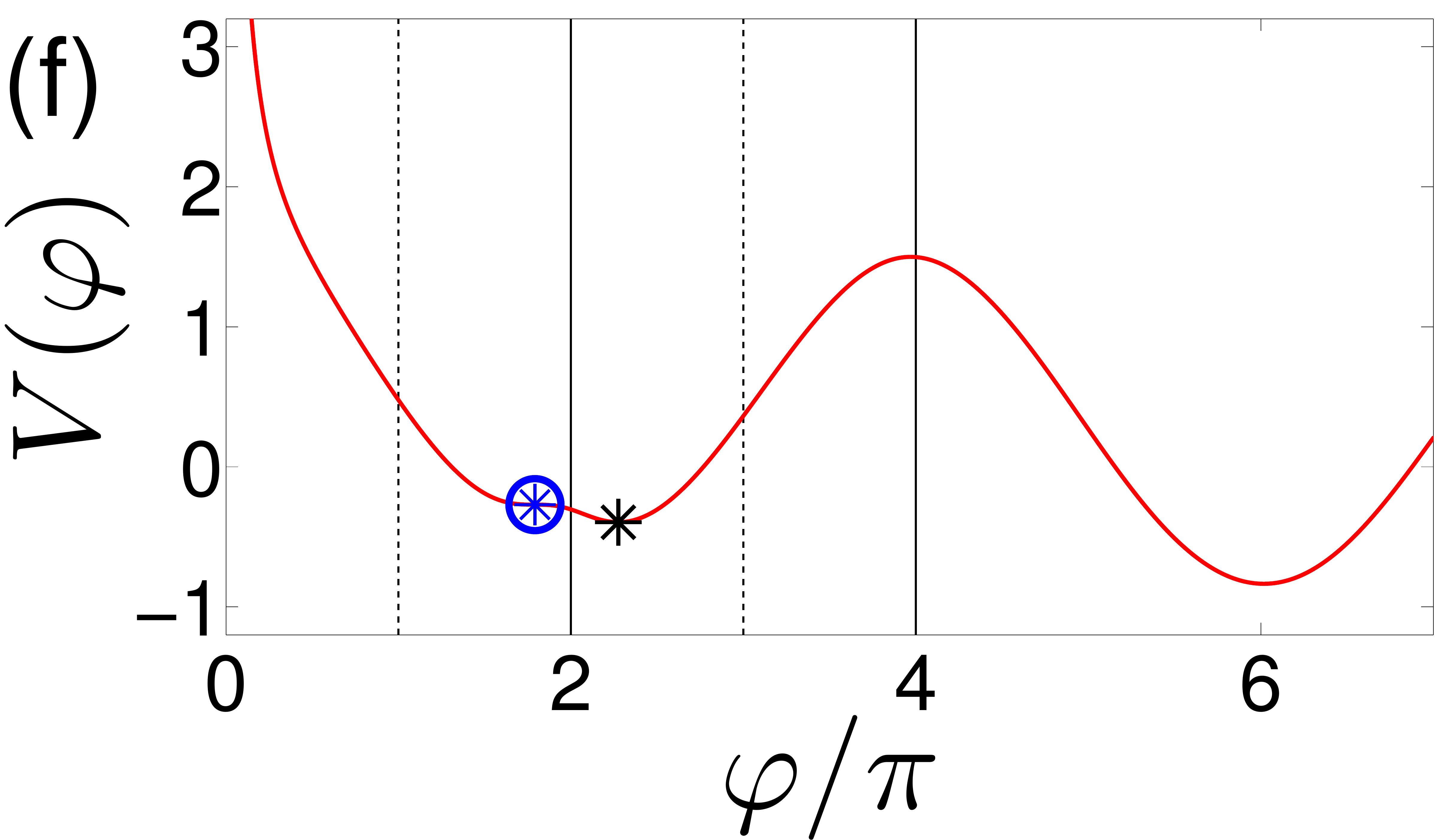}
\caption{\label{fig:SaddlenodeFigures} 
(Color online) Saddle-node bifurcation in configuration cells of type B, $r=1$, $b=0.2$. 
(a) Equilibria below the critical force, $F_x < F_x^{\text{SN,}1}$. Asterisk and circle markers indicate the particle positions in the minimum and saddle configuration, respectively, arrows show the directions into which these move with increasing $F_x$
and eventually meet, annihilating the fixed points. (b) Bifurcation diagram showing the relative coordinate $\varphi$ of the corresponding equilibrium configurations as in (a) versus $F_x$. Their COM coordinate remains unchanged with increasing $F_x$.
(c,e) Color-encoded profiles of the potential landscape $V(\varphi,\Phi)$ for $F_x < F_x^{\text{SN,}1}$ and $F_x \approx F_x^{\text{SN,}1}$, respectively, with locations of the fixed points indicated by the same markers as in (a) 
and (d,f) slices along $\Phi=\frac{3\pi}{2}$. The minimum in the type-A cell of order $r=2$ is also included here, indicated by a black asterisk.}
\end{figure}
Fig.~\ref{fig:SaddlenodeFigures}(a) shows the three-dimensional equilibrium configurations of the saddle and minimum configurations, respectively. 
The saddle configuration is characterized by both charges sitting on the side of the helix that faces away from the force, thus minimizing the field-induced potential energy, with a separation of around one winding between them. It changes only weakly with $F_x$.
In contrast, the minimum configuration has the same COM coordinate, but a separation of only slightly more than half a winding at small $F_x$. As $F_x$ is increased, the charges are further pushed towards the far edge of the helix
by the force which increases their relative separation. Eventually, the minimum and saddle equilibria become identical and annihilate, see the bifurcation diagram in Fig.~\ref{fig:SaddlenodeFigures}(b).
Figs.~\ref{fig:SaddlenodeFigures}(c-f) illustrate the corresponding changes in the potential landscape $V(\varphi,\Phi)$.
Note in Fig.~\ref{fig:SaddlenodeFigures}(e) how close the annihilation point (at which the saddle-node bifurcation occurs in the $r=1$ B cell) is to the minimum of the $r=2$ A cell. 
The saddle-node annihilation can be viewed as a merging of the potential wells around the two initial minima into one.
This observation forms the basis of the state transfer applications to be discussed in the next section.

\begin{figure}[ht!]
\centering
\includegraphics[height=0.2\textwidth]{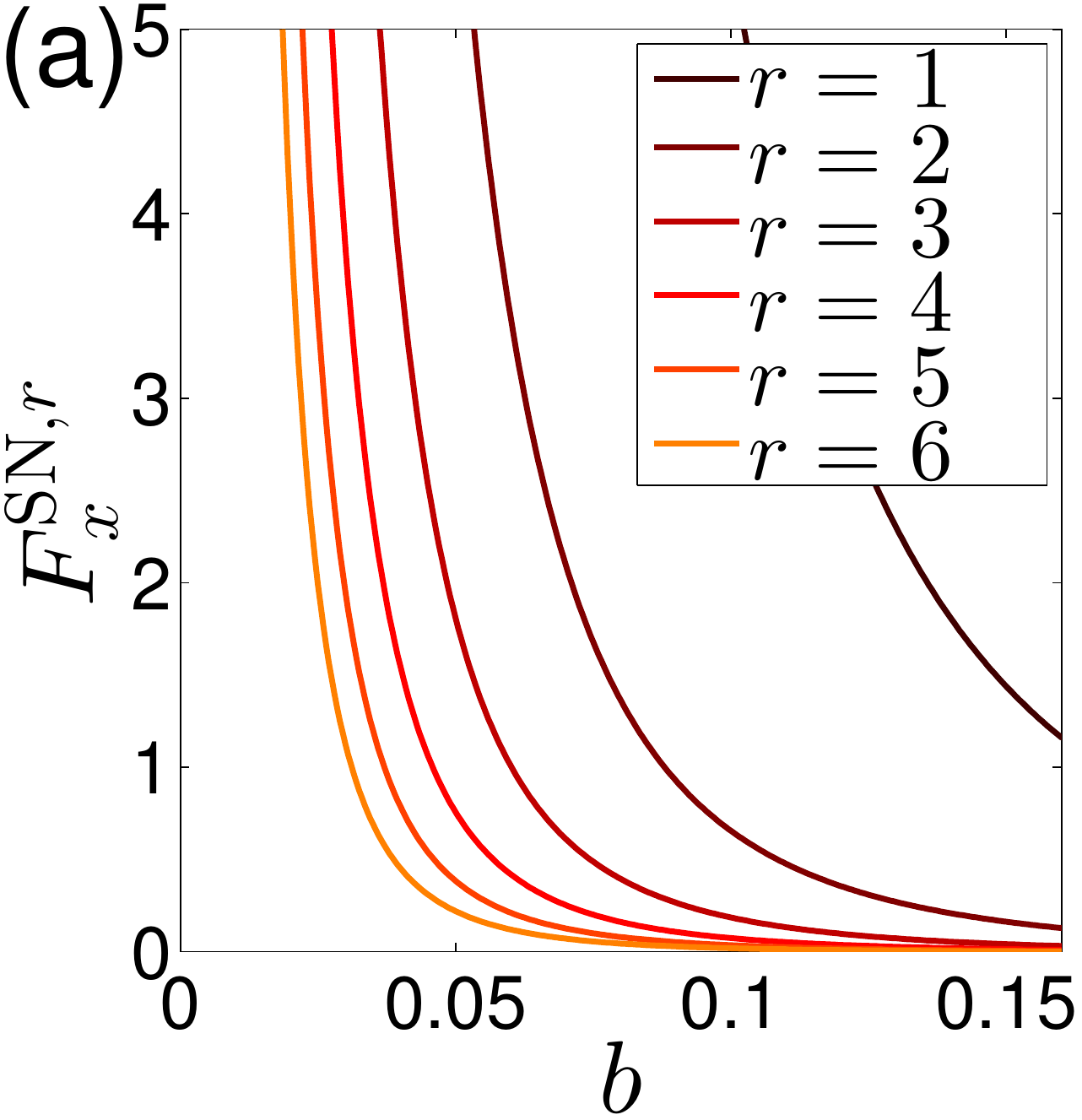} \hspace{1mm}
\includegraphics[height=0.2\textwidth]{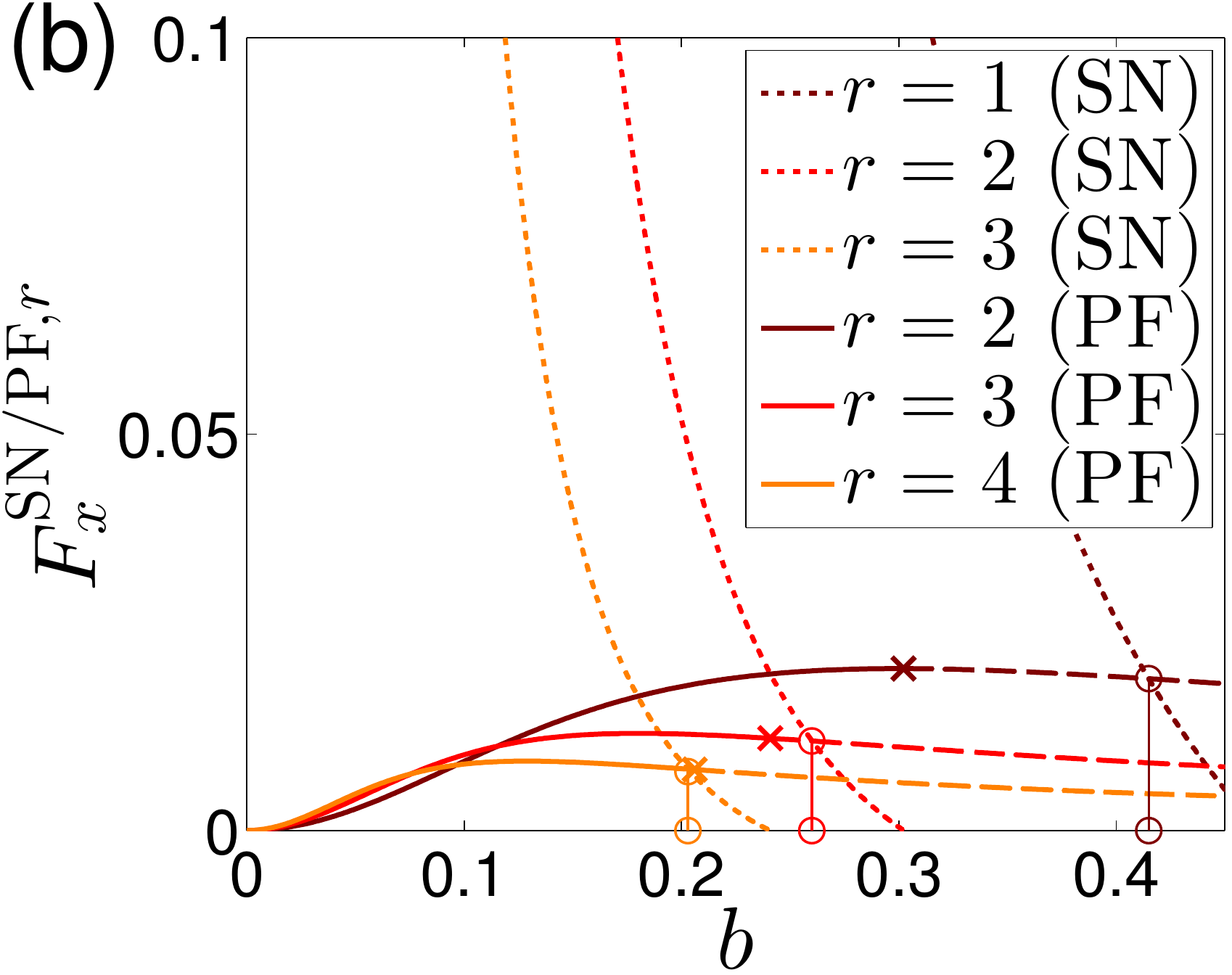}
\caption{\label{fig:FSN} 
(Color online) (a) Critical force at which the respective saddle-node bifurcation takes place in the type-B configuration space cells as a function of $b$ and the order $r$ of the cell.
(b) Comparison of the critical force for the saddle-node bifurcation in type-B cells of order $r$ to that of the pitchfork bifurcation in type-A cells of order $r+1$. 
For values of $b$ for which the former (dotted lines) is larger than the latter (solid/dashed lines as in Fig.~\ref{fig:F_zigzag}), i.e. left of the crossings indicated by circular markers, a force-driven configuration transfer as described in Section \ref{sec:transfer} is possible.}
\end{figure}

Finally, let us comment on the critical force $F_x^{\text{SN,}r}$ at which the saddle-node annihilation occurs. 
This depends both on the geometry (via the pitch parameter $b$) and the order $r$ of the configuration space cell under consideration.
In contrast to the above pitchfork bifurcations, in this case there is no closed expression for the critical force as a function of $b$ and $r$.
Fig.~\ref{fig:FSN}(a) comprises our numerical results for $F_x^{\text{SN,}r}$, which indicate a monotonic decay both with $b$ and $r$. 
Fig.~\ref{fig:FSN}(b) provides a comparison to the corresponding $F_x^{\text{PF,}r+1}$ as will be relevant for the discussion in Sec.~\ref{sec:transfer}.

\section{Applications to state transfer}
\label{sec:transfer}
\begin{figure}[ht!]
\centering
\includegraphics[width=0.49\textwidth]{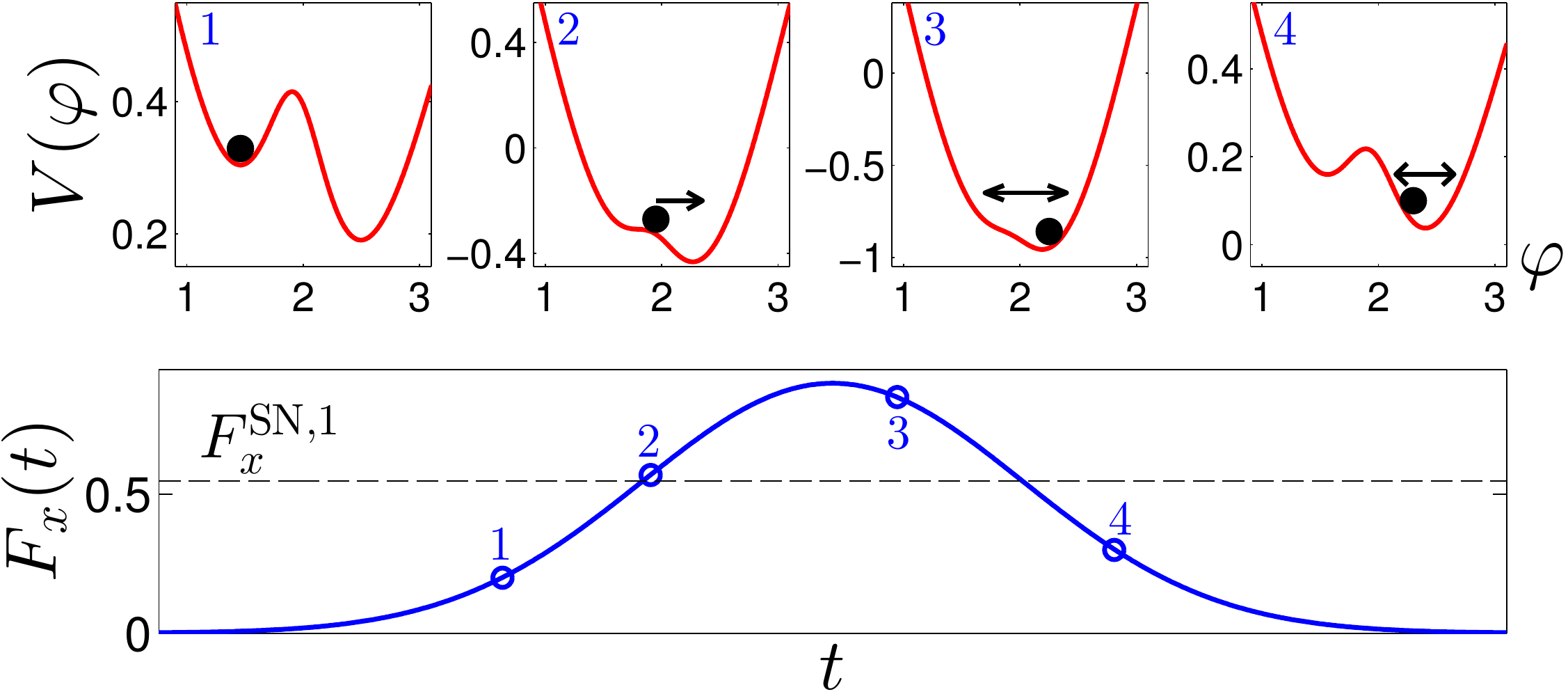}
\caption{\label{fig:transportsketch} 
(Color online) Transfer between minimum energy configurations by tuning the external force as a function of time: 
The bottom panel shows a possible $F_x(t)$ protocol, crossing the critical force $F_x^{\text{SN,}1}$ twice.
The top panels show slices through the instantaneous potential landscapes $V(\varphi)$ (at $\Phi=\frac{3}{2}\pi$ fixed and $b=0.2$) at the marked times and indicate the dynamics of the effective particle (black circle),
which initially resides in the minimum of an $r=1$ type-B cell. As $F_x$ is slowly increased, 
the particle adiabatically follows this minimum until the critical value for the saddle-node bifurcation in this cell is reached. 
Then the corresponding minimum vanishes and the effective particle starts to roll towards the close-lying minimum in the neighboring type-A cell (of order $r=2$ here). 
Slowly tuning $F_x$ back below the critical value, the $r=1$ minimum is restored but the particle still oscillates in the vicinity of the $r=2$ minimum. }
\end{figure}

Combining the above insights into the bifurcations of the fixed-point landscape makes it possible to devise protocols for controlled transfer of the two-particle system between equilibria of different order $r$ (i.e. of different separation between the charges). The key idea is to apply an external force $F_x(t)$ with a suitable time-dependence. For example, starting out in the minimum of a type-B cell of order $r$, one can (near-adiabatically) increase $F_x$ until it crosses $F_x^{\text{SN,}r}$, such that this minimum is annihilated. In the subsequent dynamics, the system moves away from the former equilibrium position which can be visualized as the motion of an effective particle in the two-dimensional $V(\varphi,\Phi)$ potential, having different effective masses in the directions of $\varphi$ and $\Phi$, cf. Eq.~(\ref{eq:Lagrangiancom}) \cite{InhomogeneousHelix}. By symmetry, this effective particle will move along the $\varphi$-axis, and it will get trapped in the minimum of the neighboring type-A cell 
of order $r+1$ (if this minimum already exists). This is schematically shown in Fig.~\ref{fig:transportsketch}. In other words, the saddle-node bifurcation can be viewed as a merging of the type-B minimum of order $r$ and the type-A minimum of order $r+1$ into a single joint potential well in which the effective particle then oscillates. 
Slowly ramping down $F_x$ below $F_x^{\text{SN,}r}$ again, the saddle-node annihilation is reversed and the order-$r$ minimum and saddle are re-established, but the system remains bound 
in the potential well of the order $r+1$ minimum. In this way, transfer between equilibrium configurations of different $\varphi$, increasing the separation between the charges, has been achieved.
We have checked the reliability of this transfer protocol in direct numerical simulations, confirming in particular its robustness against variations in the $F_x(t)$ pulse shape.

A necessary condition for this transfer scheme to be applicable is that the minimum in the A-cell of order $r+1$ forms at a smaller $F_x$ than the critical force for the saddle-node annihilation in the B-cell of order $r$, i.e. that
$F_x^{\text{SN,}r} > F_x^{\text{PF,}r+1}$.
As can be seen in Fig.~\ref{fig:FSN}(b) this is fulfilled for a large range of values of $b$ and $r$, particularly for $b$ and/or $r$ at which the involved potential wells 
become reasonably deep and thus arguably are most relevant for applications.
In cases in which $F_x^{\text{SN,}r} > F_x^{\text{PF,}r+1}$ does not hold, annihilating the minimum in the B-cell of order $r$ induces time evolution towards the saddle in the A-cell of order $r+1$,
but due to the lack of stability with respect to $\Phi$ the system is prone to perturbations in the center-of-mass coordinate.

\begin{figure}[ht!]
\begin{center}
\includegraphics[width=0.23\textwidth]{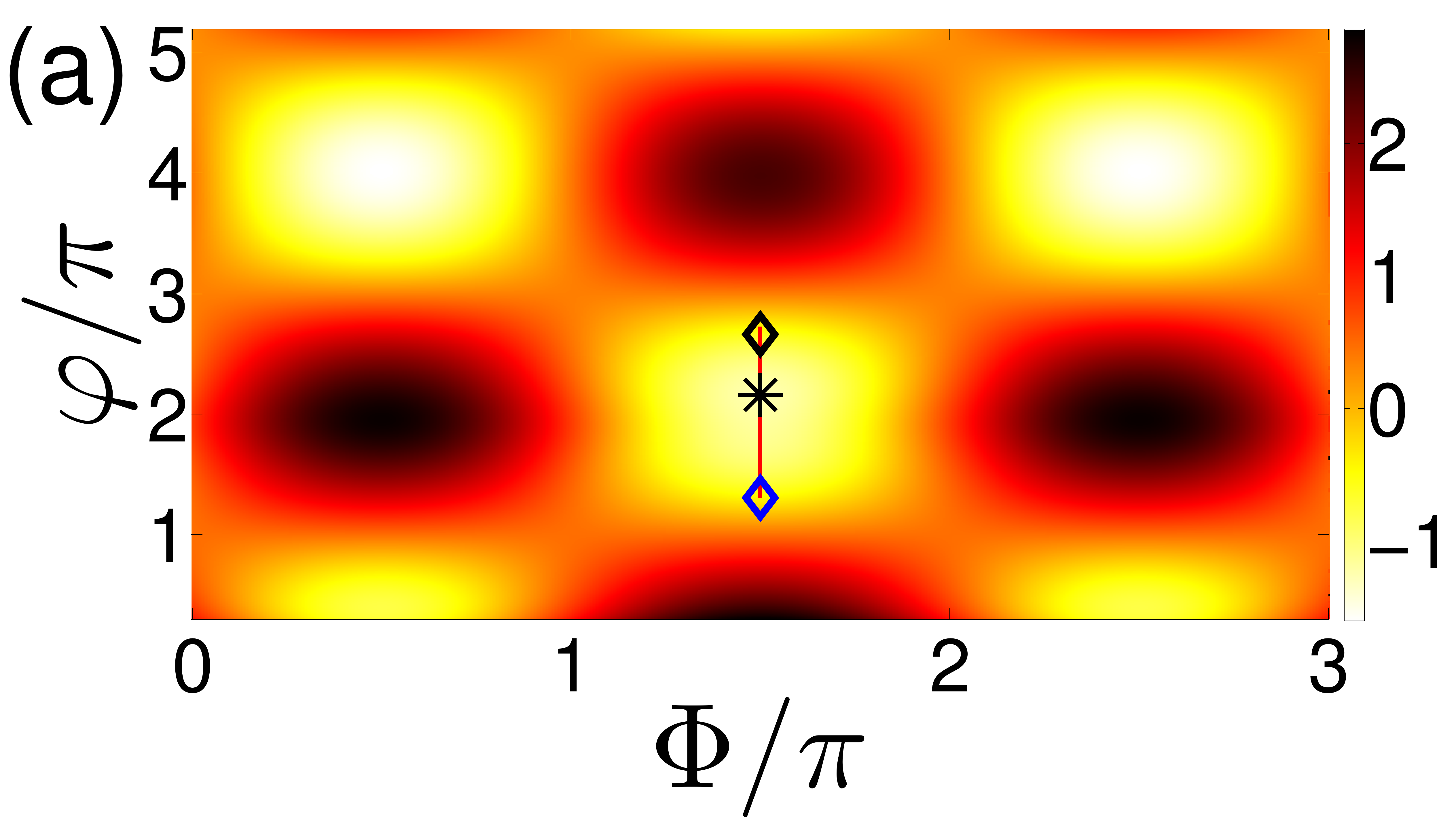}
\includegraphics[width=0.23\textwidth]{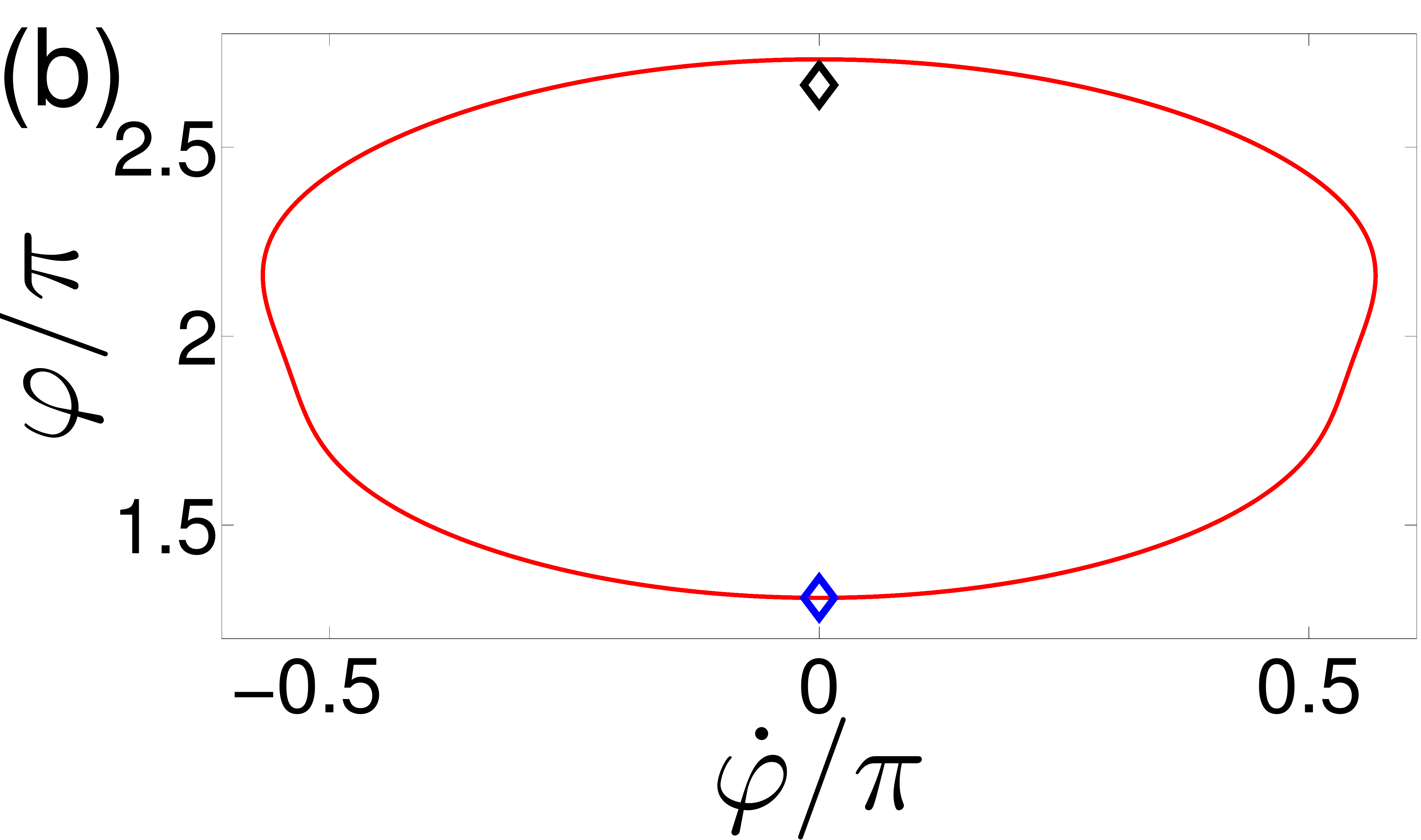}
\includegraphics[width=0.23\textwidth]{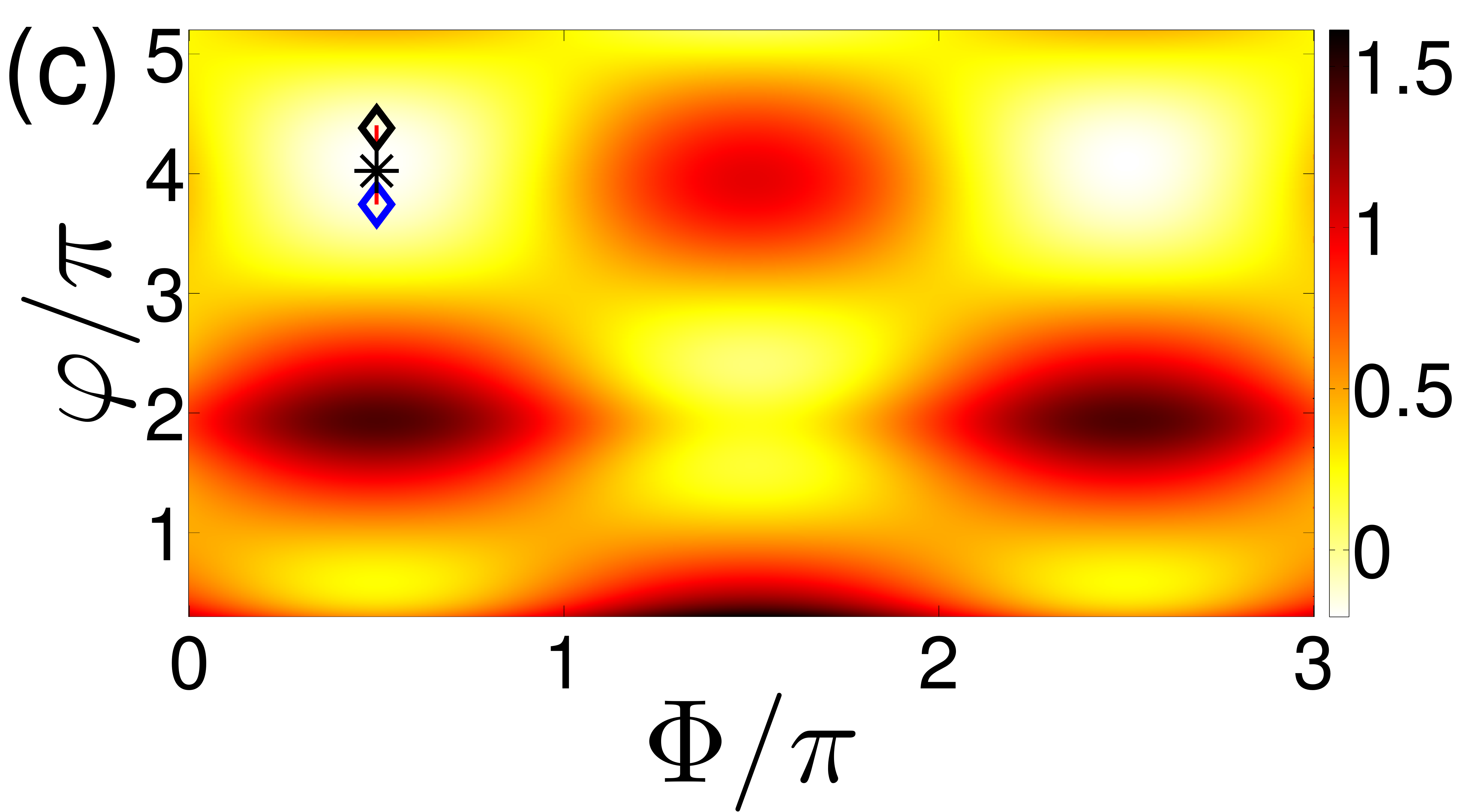}
\includegraphics[width=0.23\textwidth]{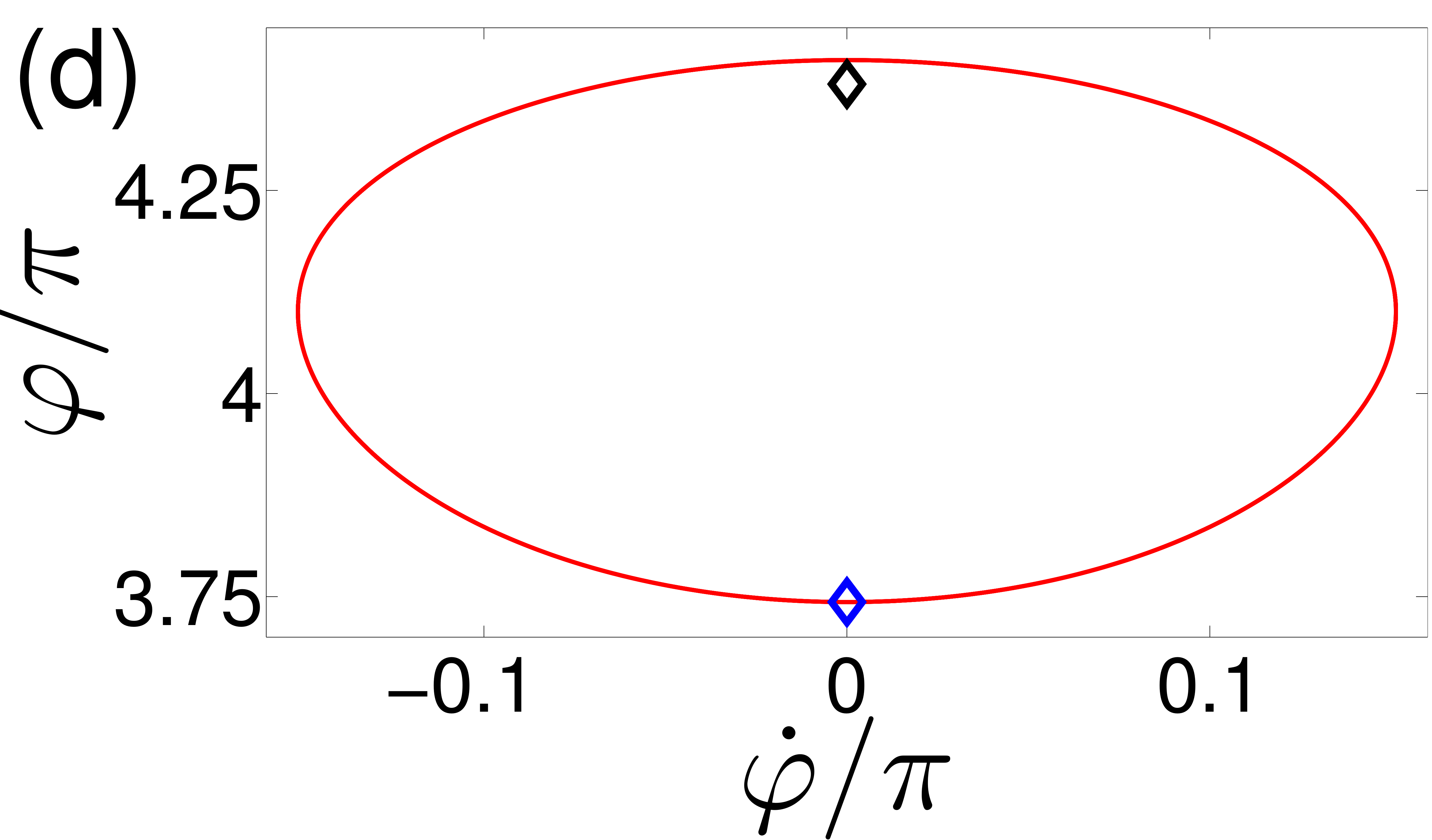}
\caption{\label{fig:Quench}
(Color online) (a) Color-encoded profile of the potential landscape $V(\varphi,\Phi)$ at $F_x'=1$, $b=0.2$. The diamond markers represent the $r=1$ (initial state) and $r=2$ (target state) minima before the quench ($F_{x}=0.1$).
As the force is switched to $F_x'$, the effective particle oscillates around the new minimum indicated by the asterisk and
follows the oscillatory trajectory indicated as a red line whose turning point lies close to the target state.
This is also seen in the phase space plot of the trajectory in (b). For these parameters, a half period of oscillation takes around $t=3.82$ and choosing a delay time $\Delta t$ for switching back to the initial $F_x$ close to this will result in the
system being trapped near the target minimum. (c,d) The quench-based transfer scheme also applies for transfer between the $r=2$ and $r=3$ minima, $F_x'=0.3$ in this example.}
\end{center}
\end{figure}

The simple transfer protocol described so far is largely insensitive to the details of the force profile $F_x(t)$ as long as it slowly crosses $F_x^{\text{SN,}r}$ twice as required. 
However, it is by construction limited to transferring the system from the B-cell minimum that is annihilated in the saddle-node bifurcation to the A-cell minimum at increased $\varphi$.
Transfer in both directions is possible by using a more rapid variation of $F_x(t)$, temporarily driving the system further away from equilibrium.
For instance, we consider sudden quenches from an initial $F_x$ to a larger value $F_x'$, holding it for a delay time $\Delta t$ and quenching back to $F_x$.
If the initial $F_x < F_x^{\text{SN,}r}$ and the intermediate $F_x' > F_x^{\text{SN,}r}$, this corresponds to a sudden switch from a two-well potential (as in panel 1 of Fig.~\ref{fig:transportsketch}) 
to a single-well potential (as in panel 3). After the quench to $F_x'$, the effective particle will thus perform oscillations in the joint well and depending on the time $\Delta t$ 
at which one switches back to $F_x$ it may end up trapped in either of the two wells.
Evidently this transfer between the minima works in both directions but requires more fine-tuned choices of $F_x'$ and $\Delta t$.
Fig.~\ref{fig:Quench}(a) illustrates the dynamics triggered by a sudden quench to a force $F_x'>F_x^{\text{SN,}r}$. The effective particle starts out at the position of one minimum of the pre-quench potential at $F_x$ 
(these former minima are indicated by diamond markers) and oscillates along $\varphi$ through the joint potential well. For the chosen value of $F_x'$, the turning
point of the trajectory lies close to the position of the desired other pre-quench minimum (diamond marker), see also the phase space plot in Fig.~\ref{fig:Quench}(b). 
Switching back from $F_x'$ to $F_x$ after a suitable delay time $\Delta t$ adapted to the oscillation period will thus leave the system close to the desired target minimum with little excess energy.

\begin{figure}[ht!]
\centering
\includegraphics[width=0.48\textwidth]{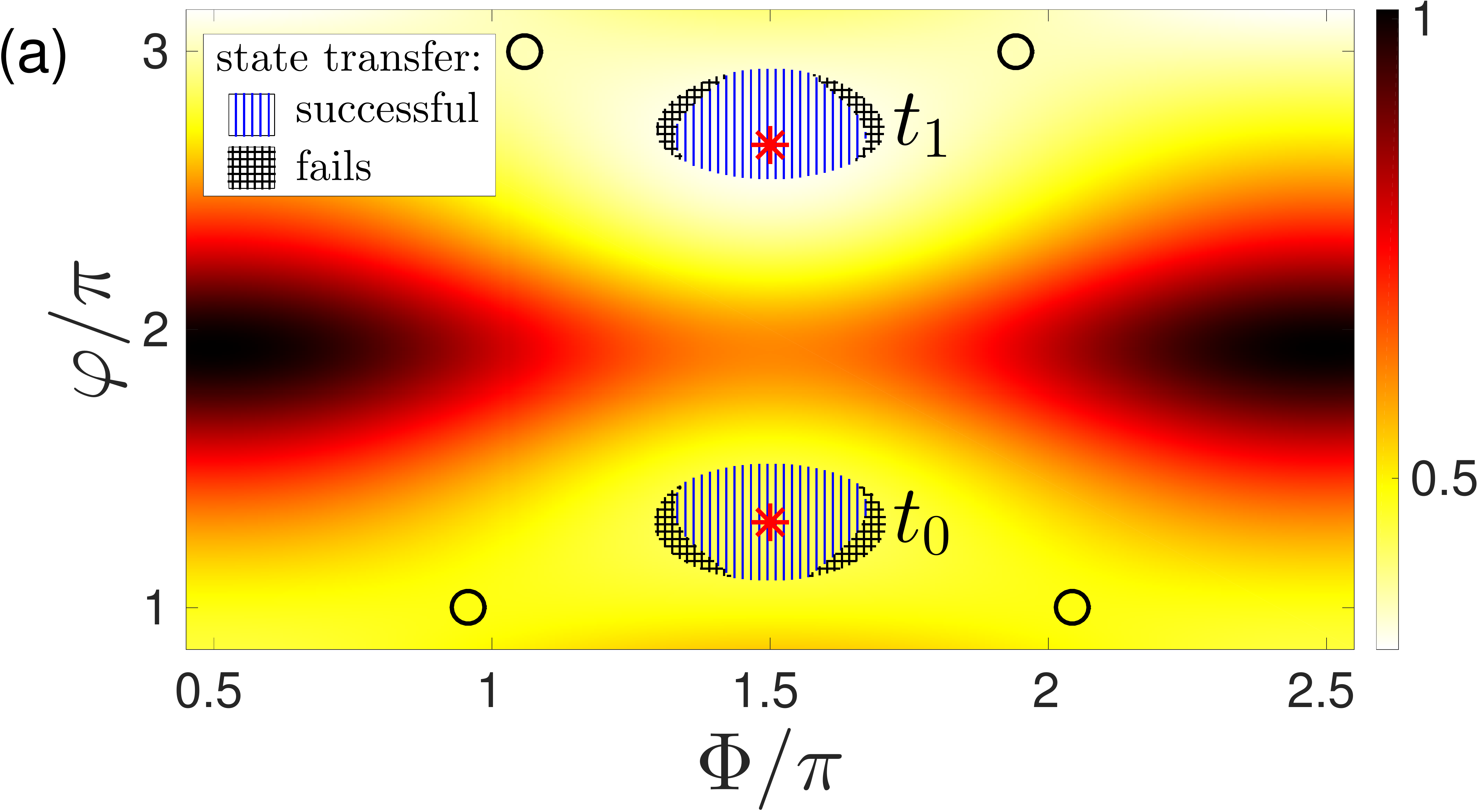}
\includegraphics[width=0.23\textwidth]{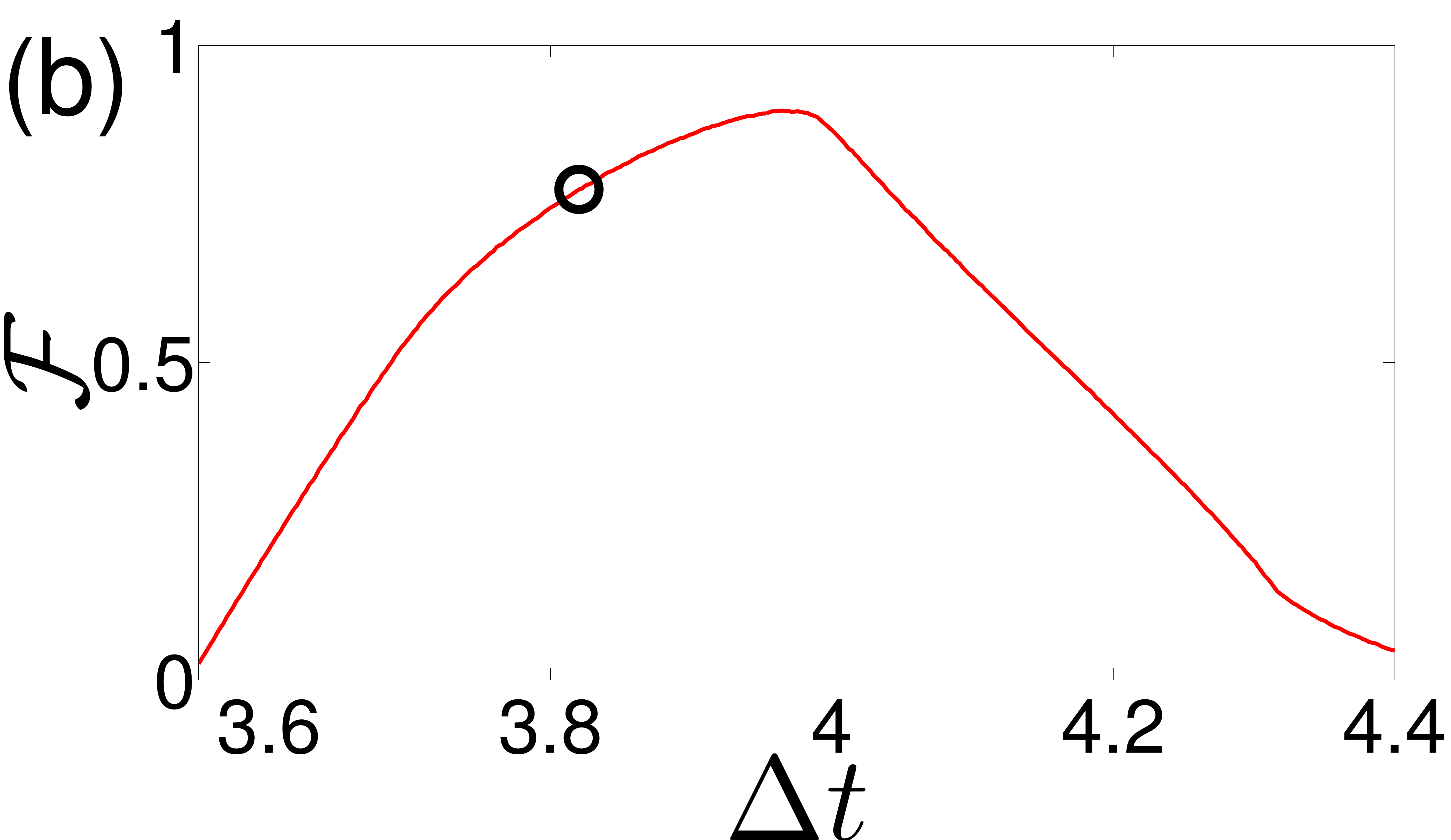}
\includegraphics[width=0.23\textwidth]{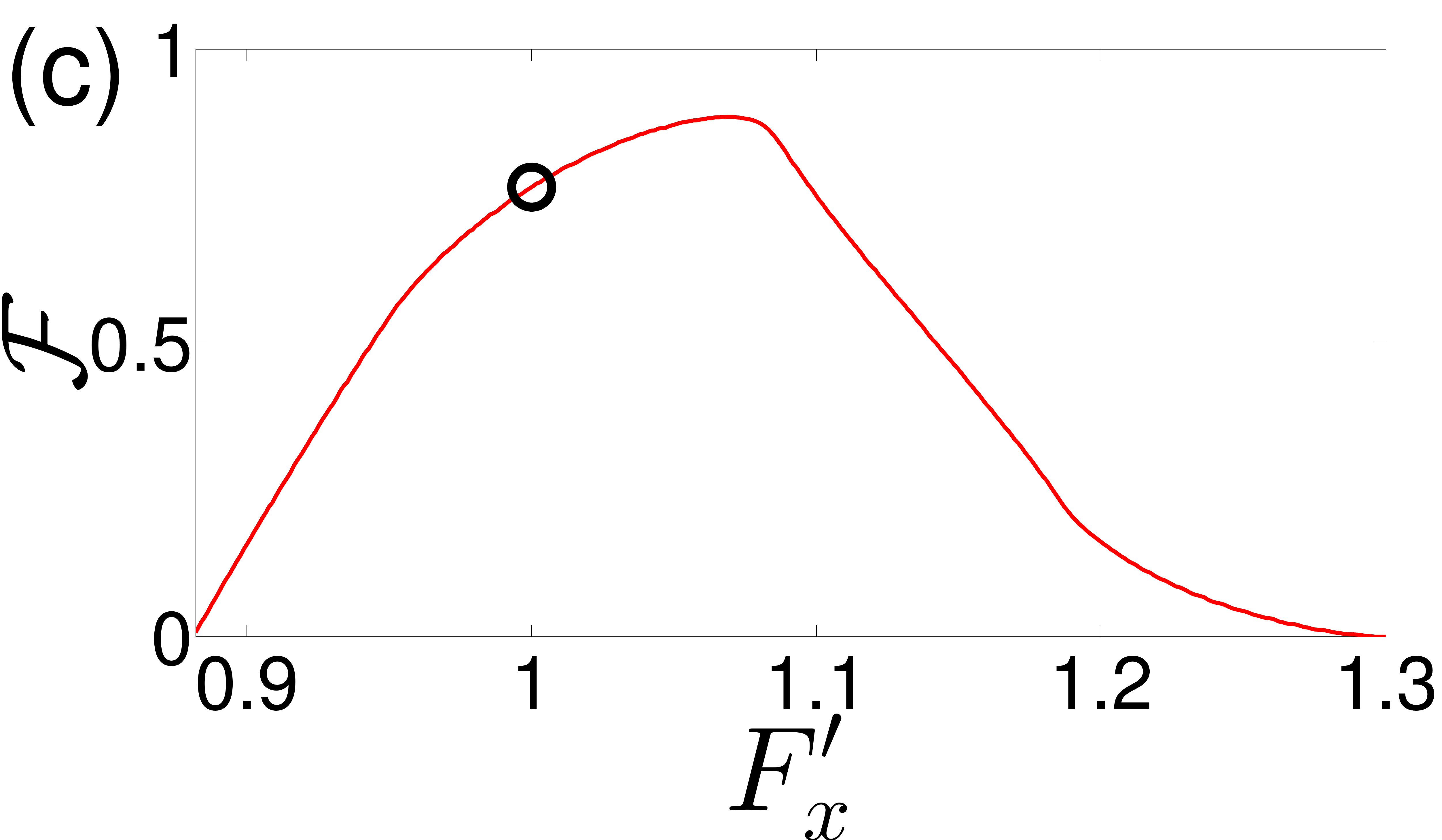}
\caption{\label{fig:InitialConditions}
(Color online) (a) Robustness of the quench-based transfer scheme: Colors encode the $V(\varphi,\Phi)$ potential landscape for $F_x=0.1$, $b=0.2$. 
At $t_0$ an ensemble of effective particles is initialized in a neighborhood of the $r=1$ minimum (lower red asterisk). 
After a quench to $F_x'=1$ each of these particles is propagated for $\Delta t=t_1-t_0=3.82$ such that the final positions lie near the $r=2$ minimum (upper red asterisk).
For each trajectory it is then evaluated if, when quenching back to $F_x=0.1$, it would end up energetically trapped in the $r=2$ minimum (blue vertical hatching, transfer successful) or untrapped (black cross hatching, transfer failed).
(b,c) Fraction of successfully transferred trajectories $\mathcal F$ as a function of the delay time $\Delta t$ (at fixed $F_x'=1$) and $F_x'$ (at fixed $\Delta t=3.82$), respectively. Circular markers indicate the parameters used in (a).}
\end{figure}

The quench-based transfer protocol is relatively stable in terms of varying the initial conditions. 
Quantifying this for a special case, let us analyze the transfer from $r=1$ to $r=2$ at $b=0.2$, $F_{x}=0.1$, i.e. as in Fig.~\ref{fig:Quench}(a,b).
Instead of just a single trajectory as in that figure, we consider now an ensemble of effective particles, initialized in a circular neighborhood of the $r=1$ pre-quench minimum at $F_x$. 
We quench to a force $F_x'>F_x^{\text{SN,}r}$ and back to $F_x$ after a delay time $\Delta t$ and then evaluate for each trajectory whether it ends up being energetically trapped near the $r=2$ target minimum or not.
For one combination of $F_x'$ and $\Delta t$ the result is shown in Fig.~\ref{fig:InitialConditions}(a), demonstrating that although the ensemble is initially spread relatively widely, a large fraction of trajectories indeed ends
up trapped near the target state.
As a quantitative measure of this, we consider the fraction $\mathcal F$ of successfully transferred trajectories in the ensemble.
Even for the rather widespread ensembles we use, this adopts maximum values of around $90\%$ and upon deviating from the optimal parameters $F_x'$ and $\Delta t$ it still remains substantially large, see Figs.~\ref{fig:InitialConditions}(b,c).

Such a quench-based protocol thus allows robust bi-directional configuration transfer between the stable equilibria of type-B, order $r$ and type-A, order $r+1$ without affecting the center of mass.
To complete the transfer toolbox, it is desirable to have a means of moving the center-of-mass coordinate in a controlled way. 
As seen above, $\Phi$ can in principle be accelerated and decelerated by a force component $F_z$ along the helix axis. 
Alternatively, one can employ a force that is purely perpendicular to the helix axis and rotate it, i.e. in Cartesian coordinates $\mathbf{F}= F (\cos \alpha, \sin \alpha, 0)$.
Then the force-induced contribution to the potential generalizes to 
\begin{equation}
 V_{F}(\varphi,\Phi)= - 2F \sin \left( \Phi + \alpha \right) \cos \frac{\varphi}{2},
\end{equation}
while the $\varphi$-dependent interaction potential is, of course, unaffected. This means that adiabatic tuning of the rotation angle $\alpha$ leads
to transport of the center-of-mass coordinate without changing the relative position of the charges.
Specifically, one can start with a force along the $x$-axis (as done above) and rotate it by a multiple of $2\pi$. This admits a simple transfer 
between type-A (or type-B) equilibria of different $k$ (i.e. horizontally in Fig.~\ref{fig:EquilibriaLandscape}).

Transfer from a type-A to a type-B cell of the same order $r$ is also possible, for instance with the following protocol: 
Start in a type-B minimum with a force along the $x$-axis and rotate by $\alpha=\pi$. Then the COM of the charge pair has already been transferred to the desired position. 
After slowly decreasing $F_{x}$ to almost $F^{\text{PF},r}_x$ (such that the minimum is not lost crossing the pitchfork bifurcation point) 
and then quenching back to $\alpha=0$, the configuration will be close to the desired type-B equilibrium and oscillate around it.
An intermediate reduction of the absolute value of the force is helpful here in order to suppress the potential gradient along $\Phi$, that may otherwise induce undesired COM motion after the second quench.
It also serves to bring the $\varphi$-positions of the pre-quench and target states closer together and thus reduce the excess energy.
Subsequently, $F_{x}$ can be increased back to its initial value to complete the transfer.
The reverse transfer from a B-cell to an A-cell of identical $r$ can essentially proceed along the same lines; since one does not start in a type-A minimum here,
one can even drop the restriction $F_x(t) > F^{\text{PF},r}_x$ for all times.

\section{Conclusions}
\label{sec:con}
We have investigated the local equilibria and classical dynamics of two charged particles confined to a helix and subject to a constant external electric field.
In the absence of the field the effective two-body interaction potential exhibits multiple minima depending on the geometry parameters. The existence of a finite transverse 
electric field induces a coupling of the center-of-mass to the relative coordinate and thus alters the total potential landscape by localizing the center-of-mass coordinate i.e.
favoring particular values of it. By increasing the external field amplitude the potential landscape keeps altering and various critical points (saddles, minima or maxima) 
emerge or annihilate through different bifurcation scenarios, signifying the very rich nonlinear behavior of this system.

Among the observed effects there exists a merging of two potential wells corresponding to different relative configurations into one with increasing the magnitude of the external force.
By providing the force with a certain time dependence, we succeed in transferring the particles from one state into another in which the charges are separated 
by a different interparticle distance. Apart from these transfers being in general robust with respect to the particular type of the time dependence, different 
quench protocols can be used to achieve transfer between various states including also states with different values of the center-of-mass coordinate. Therefore
by choosing a suitable protocol even a transfer between arbitrary states is possible, at the cost of acquiring in general much excess energy which should be
somehow drained from the particles (e.g. through friction) in order to retain control over the transfer.
Such transfer mechanisms are reminiscent of the field-driven charge pumping investigated recently  in \cite{Topological1} for a tight-binding system with a helical structure
argued to be a suitable model for helical molecules.

Concerning the experimental realization of such a setup certain advances have been made apart from the field of nanofabrication \cite{Graphene_Helix, HelixNanotubes1, HelixNanotubes2} also 
in the field of ultracold atoms \cite{Nanofibers3,LaguerreBeam2,LaguerreBeam1} where the realization of optical helical traps has recently been proposed.
From the theory side, further studies could be dedicated to investigations of the many-body analogue of the present system.
Given the variety and the wealth of effects in much simpler models which combine both an external and an interaction potential, an example being the Frenkel-Kontorova model \cite{FrenkelKontorova},
it is natural to expect this to hold as well for the many-body system of helically confined charges in the presence of an external field where especially the 
effective interaction potential is already complex.


\bibliography{Sources}

\end{document}